\documentclass[twocolumn,preprintnumbers,elsart]{revtex4}
\usepackage{eurosym}
\usepackage{makeidx}
\usepackage{amssymb}
\usepackage{amsmath}
\usepackage{mathrsfs}
\usepackage{graphicx}
\usepackage{dcolumn}
\usepackage{bm}
\usepackage[center]{subfigure}
\usepackage{color}
\usepackage{indentfirst}
\usepackage{array}
\usepackage{booktabs}
\usepackage{multirow}
\usepackage{upgreek}
\usepackage{float}

\begin{document}

\title{Formation and dynamics of self-bound droplets in dipolar molecular condensate}
\author{Xinyi Tang$^{1}$}
\thanks{These authors contributed equally to this work.}
\author{Tianmiao Zhang$^{1}$}
\thanks{These authors contributed equally to this work.}
\author{Zibin Zhao$^{1}$}
\author{Guilong Li$^{2}$}
\author{Zhaopin Chen$^{3}$}
\author{Bin Liu$^{1,4}$}
\email{binliu@fosu.edu.cn}
\author{Boris A. Malomed$^{1,5}$ }
\author{Yongyao Li$^{1,4}$ }
\email{yongyaoli@gmail.com}
\affiliation{$^1$School of Physics and Optoelectronic Engineering, Foshan University,
Foshan 528000, China\\
$^2$College of Engineering and Applied Sciences, National Laboratory of
Solid State Microstructures, Nanjing University, Nanjing 210023, China\\
$^3$Physics Department and Solid-State Institute, Technion, Haifa 32000,
Israel \\
$^4$Guangdong-Hong Kong-Macao Joint Laboratory for Intelligent Micro-Nano
Optoelectronic Technology, Foshan University, Foshan 528225, China \\
$^5$Department of Physical Electronics, School of Electrical Engineering,
Faculty of Engineering, Tel Aviv University, Tel Aviv 69978, Israel }

\begin{abstract}
We study self-bound quantum droplets in the regime dominated by microwave-induced non-axisymmetric dipole-dipole interactions, using the extended Gross-Pitaevskii equation with the Lee-Huang-Yang corrections. We identify the existence region through numerical simulations and employ an anisotropic Gaussian-super-Gaussian variational ans\"{a}tz to capture the characteristic density profile of the droplets, with a Gaussian profile along the narrow $x$ direction and super-Gaussian profiles in the extended $(y,z)$ plane. Within this variational framework, we characterize the self-binding, spatial localization, and density-compression properties of the droplets and find good agreement between the variational predictions and the numerical results. Collisions between droplets moving along different directions reveal a strong directional dependence, with outcomes ranging from quasi-elastic rebound and merger to fragmentation. In addition, we explore the rotational dynamics of a single self-bound droplet about all three Cartesian axes, revealing rich and controllable three-dimensional rotational dynamics. Together, these results demonstrate how non-axisymmetric dipolar interactions provide versatile means for controlling the translational, collisional, and rotational dynamics of self-bound quantum droplets.

\end{abstract}
\maketitle

\section{Introduction}

Ultracold dipolar molecules~\cite{Krem2008, Baranov2008, Ni2008, Moses2017}
are emerging as a promising testbed for many-body physics \cite{Lahaye2009,
Baranov2012}, quantum simulations~\cite{Micheli2006, Carr2009, Altman2021},
and quantum data processing~\cite{Baillie2002, Rabl2006}. A major challenge
in these settings arises from severe collisional losses due to short-range
molecular encounters~\cite{Ospelkaus2010, Quemener2012, Bohn2017, XY2018,
Bause2023,Gorshkov2008}.~Recent advances in microwave shielding have helped
to significantly suppress these losses, enabling the creation of long-lived
molecular Bose-Einstein condensates~(BECs) \cite{Karman2018,
Lassabliere2018, Anderegg2021, Schindewolf2022, Chen2023, Lin2023,
Bigagli2023, Deng2023, Chen2024, Dutta2025}.

Building on this progress, it has been demonstrated that dual-frequency
microwave-field shielding~\cite{Karman2025, Bigagli2024, Yuan2025, Shi2025,
Deng2025,Arnone2026,Jin2025,Li2024} enables highly tunable strong long-range
dipole-dipole interactions (DDIs), thus providing a control level beyond
that achievable in BEC of magnetic atoms \cite{FerrierBarbut2016,
Schmitt2016, Kadau2016, Boudjemaa2018, Semeghini2018, Cabrera2018,
Chomaz2022}.
In the dual-field dressing scheme, the ellipticity of the
$\sigma$ field introduces an non-axisymmetric component of the DDI, proportional to
$\sim \sin^2\theta\cos2\phi$, which breaks the axial symmetry and makes the interaction
direction-dependent in the transverse plane.~\cite{Xu2025,Karman2025}
The linearly polarized $\pi$  field provides an additional control channel for
the usual axially symmetric component.
By tuning the microwave amplitudes and detunings along a physically allowed dressing trajectory,
the axially symmetric component can be compensated while a non-axisymmetric interaction remains.~\cite{Ho2026}

The recent experimental realization \cite{experiment,Langen2025} of
self-bound molecular quantum droplets (QDs) in the BEC of NaCs
molecules~with strong DDI \cite{langen2025} has put forward a new setup for
the work with self-trapped states in the polar-molecule BEC \cite%
{Schmidt2022}, similar to QDs in atomic BECs \cite{Petrov2015,Petrov2016,
Bisset2016, Lima2011, FerrierBarbut2016Review,Guo2020}. The recent work by Baillie \cite{Baillie} established a comprehensive framework for microwave-dressed dipolar quantum droplets, systematically mapping their ground-state (GS) structures and properties across the two interaction parameters, $\epsilon_{dd}^{(0)}$ and $\epsilon_{dd}^{(2)}$, using Gaussian variational and extended Gross-Pitaevskii (GPE) approaches. Here, $\epsilon_{dd}^{(0)}$ and $\epsilon_{dd}^{(2)}$ characterize the relative strengths of the axisymmetric and non-axisymmetric DDI, respectively. Building on this framework, we concentrate on the purely non-axisymmetric limit, $\epsilon_{dd}^{(0)}=0$ and $\epsilon_{dd}^{(2)}\neq0$, where the droplets acquire strongly anisotropic, flattened density profiles. In this regime, we refine the variational description by adopting an anisotropic Gaussian-super-Gaussian ans\"{a}tz and extend the analysis beyond equilibrium properties to the dynamics of self-bound droplets, including their directional collision behavior and rotational motion about the three Cartesian axes.

The subsequent presentation is organized as follows. Section II introduces the
theoretical model. Section III analyzes the ground-state properties using the
 Gaussian-super-Gaussian variational ans\"{a}tz and numerical simulations.
  Section IV addresses head-on collisions between identical QDs and the
  rotational response of a single QD. Section V summarizes the main
  results and discusses directions for future work.

\section{The model}

\begin{figure}[tbp]
\centering
\includegraphics[width=0.5\textwidth, trim=0cm 0cm 0cm 0cm, clip]{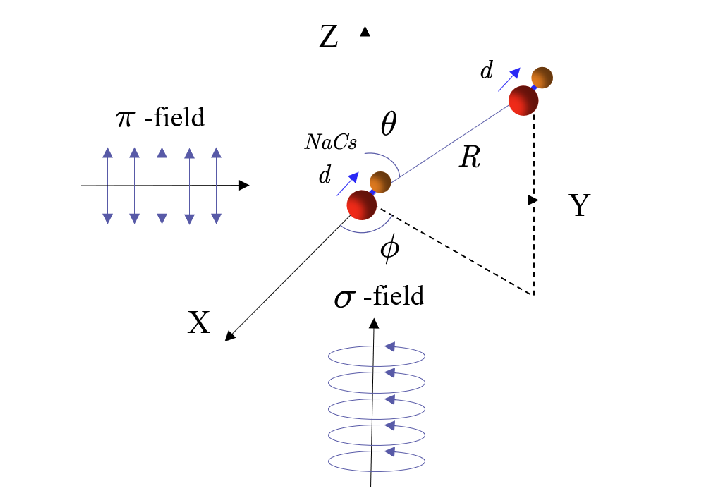}
\caption{ Schematic illustration of the control microwave fields and dipolar
interaction geometry. }
\label{fig:Fin1}
\end{figure}
We consider BECs of electric dipolar bialkali molecules~\cite{experiment},
which are dressed by two microwave fields~\cite%
{Karman2025,Zhang2026,Schindewolf2026}. These are a linearly polarized field
oscillating along the $Z$-axis (the $\pi $-field), and an elliptically
polarized one rotating in the $(X,Y)$ plane (the $\sigma $-field),
respectively, see a schematic illustration of the control fields in Fig.~\ref%
{fig:Fin1}. The interaction between the dipole molecules is controlled by a
strongly repulsive short-range hardcore contact interaction and the
long-range DDIs with the potential \cite{Karman2025,Chen2023}
\begin{equation}
U_{dd}(\mathbf{R})=\frac{3G_{s}}{4\pi R^{3}}\left[\epsilon_{dd}^{(0)}(1-3%
\cos^2\theta)+\sqrt{3}\epsilon _{dd}^{(2)}\sin ^{2}\theta \cos 2\phi\right],
\label{DDI}
\end{equation}%
where $\theta $ is the angle between $\mathbf{R}$ and $Z$-axis, and $\phi $
is the angle between the projection of $\mathbf{R}$ onto the $(X,Y)$ plane
and the $X$-axis, $m$ and $\hbar $ being the mass of the molecule and Planck
constant, respectively. Coefficient $G_{s}={4\pi \hbar ^{2}a_{s}}/{m}$ is
the contact-interaction strength, with $a_{s}$ being the \textit{s}-wave
scattering length.

The first term of Eq.~\eqref{DDI} represents the usual axially symmetric DDI
between dipolar molecules, which can be tuned by the combined effects of the
two control microwaves~\cite{experiment} with a relative strength $\epsilon
_{dd}^{(0)}=a_{d0}/a_{s}$, where $a_{d0}$ is the effective DDI scattering
length. The second term of Eq.~\eqref{DDI} represents the non-axially
symmetric DDI between the dipolar molecules, which is controlled by the
ellipticity of the $\sigma $-field~\cite{experiment,Karman2025} with a
relative strength $\epsilon _{dd}^{(2)}=a_{d2}/a_{s}$, where $a_{d2}$ is the
respective DDI scattering length. Here, both $\epsilon _{dd}^{(0)}$ and $%
\epsilon _{dd}^{(2)}$ can be tuned over a wide range of positive, zero and
negative values.

The angular structure of Eq.~\eqref{DDI} can be naturally understood in
terms of the spherical harmonics $Y_{lm}(\theta ,\phi )$. In particular, the
first term is proportional to the spherical harmonic $Y_{20}(\theta ,\phi )$%
, corresponding to the $m=0$ component. This term preserves rotational
symmetry about the $z$-axis and represents the conventional dipole-dipole
interaction between polarized molecules. In contrast, the second term is
proportional to the real linear combination $Y_{2,2}(\theta ,\phi
)+Y_{2,-2}(\theta ,\phi )$, corresponding to the $m=\pm 2$ components. This
term explicitly breaks the axial symmetry and introduces a four-fold
anisotropy in the transverse $(x,y)$ plane.

The evolution of the system's wave function $\Psi $ is governed by the
extended GPE (alias eGPE)~\cite{Baillie},
\begin{equation}
\begin{gathered} i\hbar\frac{\partial}{\partial T}\Psi = \Bigg[
-\frac{\hbar^2}{2m}\nabla_{XYZ}^2 + G_s\left|\Psi\right|^2 \\+ \int
U_{\mathrm{dd}}(\mathbf{R}-\mathbf{R}^\prime)\left|\Psi(\mathbf{R}^\prime)%
\right|^2 d\mathbf{R}^\prime + \Gamma_{QF}\left|\Psi\right|^3 \Bigg]\Psi,
\end{gathered}  \label{eGPE}
\end{equation}%

Here, the effect of quantum fluctuations (QFs) is represented by the strength%
\begin{equation}
\Gamma _{\mathrm{QF}}=\frac{32G_{s}a_{s}^{3/2}}{3\sqrt{\pi }}
\text{Re}\{\mathcal{Q}_{5}(\epsilon _{dd}^{(0)},\epsilon _{dd}^{(2)})\},  \label{eq:GammaQF}
\end{equation}%
of the Lee-Huang-Yang (LHY) term \cite{Baillie}, with
\begin{equation}
\mathcal{Q}_{5}(\epsilon _{dd}^{(0)},\epsilon _{dd}^{(2)})=\int \frac{%
d\Omega _{k}}{4\pi }\left[ 1+\frac{\tilde{U}_{\mathrm{dd}}(\mathbf{k})}{G_s}\right]
^{5/2}.  \label{eq:Q5}
\end{equation}%

Here, $\Omega _{k}$ is the solid angle in $k$-space, and $\tilde{U}_{%
\mathrm{dd}}(\mathbf{k})$ is the Fourier transform of $U_{\mathrm{dd}}(%
\mathbf{R})$, is given by
\begin{equation}
\tilde{U}_{\mathrm{dd}}(\mathbf{k})
=G_s\Bigg[\epsilon_{\mathrm{dd}}^{(0)}
\left(\frac{3k_Z^2}{k^2}-1\right)
-\sqrt{3}\epsilon_{\mathrm{dd}}^{(2)}
\frac{k_X^2-k_Y^2}{k^2}\Bigg],
\label{eq:Fourier_DDI}
\end{equation}

In the numerical evaluation of this Fourier transform, a
spherical cutoff of radius $R_{c}$ is introduced to regularize the DDI
integral, which gives rise to a cutoff function $s(kR_{c})$, the full
momentum-space potential $\tilde{U}_{\mathrm{dd}}(\mathbf{k})$ being
recovered in the limit of $R_{c}\rightarrow \infty $. Further details of the
calculation of $\mathcal{Q}_{5}(\epsilon _{dd}^{(0)},\epsilon _{dd}^{(2)})$
are provided in Appendix A.

The total particle number $N$ is defined by wave function $\Psi $ as
\begin{equation}
N=\int |\Psi (\mathbf{R})|^{2}d\mathbf{R}.
\end{equation}%

By means of rescaling,
\begin{equation}
R=l_{0}{r},\quad T=t_{0}t,\quad \Psi =l_{0}^{-3/2}\psi ,\quad \frac{%
ml_{0}^{2}}{\hbar t_{0}}=1,
\end{equation}
with $\psi$ denoting the corresponding dimensionless wave
function.
To obtain self-bound stationary states, we consider solutions of the form
\begin{equation}
\psi (\mathbf{r},t)=\varphi (\mathbf{r})e^{-i\mu t},  \label{1}
\end{equation}%

where $\mu $ is the chemical potential and $\varphi (\mathbf{r})$ is a
time-independent wave function. Substituting this~into Eq. yields the stationary eGPE
\begin{equation}
\begin{gathered} \label{stationary-GPE} \mu\varphi =
-\frac{1}{2}\nabla^2_{xyz}\varphi + g_s |\varphi|^2\varphi \\+ \varphi \int
u_{\mathrm{dd}}(\mathbf r-\mathbf r') |\varphi(\mathbf r')|^2 d\mathbf r' +
\gamma_{QF} |\varphi|^3\varphi . \end{gathered}
\end{equation}

Here, the dimensionless strength of the contact interaction
is $g_s=4\pi a_s/l_0$, while the LHY coefficient is
 $\gamma _{QF}=4{g_{s}}^{5/2}\text{Re}\{\mathcal{Q}_{5}(\epsilon _{dd}^{(0)},
 \epsilon _{dd}^{(2)})\}/(3\pi ^{2})$. The dimensionless dipolar
 interaction potential is given by
 $u_{\mathrm{dd}}({\mathbf{r}})=3g_{\mathrm{s}}/(4\pi r^{3})[\epsilon_{dd}^{(0)}
 (1-3\cos^{2}\theta)+\sqrt{3}\epsilon_{dd}^{(2)}\sin^{2}\theta\cos 2\phi]$.
Obviously, the integral $\int |\psi (\mathbf{r})|^{2}d\mathbf{r}=N$.

The energy functional associated with the stationary eGPE is~\cite{Baillie}
\begin{equation}
E=E_{p}+E_{MF}+E_{LHY}+E_{\mathrm{dd}},  \label{E}
\end{equation}%
where
\begin{equation}
E_{p}=\int \left[ \frac{1}{2}|\nabla _{xyz}\varphi (\mathbf{r})|^{2}\right] d%
\mathbf{r}, \label{E_p}
\end{equation}%

\begin{equation}
E_{MF}=\int \left[ \frac{g_{s}}{2}|\varphi (\mathbf{r})|^{4}\right] d\mathbf{%
r}, \label{E_MF}
\end{equation}%

\begin{equation}
E_{LHY}=\int \left[ \frac{2}{5}\gamma _{QF}|\varphi (\mathbf{r})|^{5}\right]
d\mathbf{r}, \label{E_LHY}
\end{equation}%
\begin{equation}
E_{\mathrm{dd}}=\frac{1}{2}\int d\mathbf{r}\int d\mathbf{r}^{\prime }\,u_{%
\mathrm{dd}}(\mathbf{r}-\mathbf{r}^{\prime })|\varphi (\mathbf{r}%
)|^{2}|\varphi (\mathbf{r}^{\prime })|^{2}. \label{E_dd}
\end{equation}%
To quantify the spatial localization of the QD, we define the effective
volume\cite{Zhao2026}
\begin{equation}
V=\frac{\left( \int |\varphi (\mathbf{r})|^{2}d\mathbf{r}\right) ^{2}}{\int
|\varphi (\mathbf{r})|^{4}d\mathbf{r}}.  \label{V}
\end{equation}

In the following, all quantities are presented in dimensionless units. We
choose $l_{0}=2~\mu \mathrm{m}$, yielding the characteristic time scale $%
t_{0}=ml_{0}^{2}/\hbar \simeq 9.83~\mathrm{ms}$. Consequently, $r=1$ and $%
t=1 $ correspond to physical scales of $2~\mathrm{\mu m}$ and $9.83~\mathrm{%
ms}$, respectively. The physical density is related to the dimensionless
density through $|\Psi |^{2}=l_{0}^{-3}|\psi |^{2}$, such that $|\psi
|^{2}=1 $ corresponds to the density of $1.25\times 10^{11}~\mathrm{cm}^{-3}$.

In what follows below we set $\epsilon_{dd}^{(0)}=0$, to
focus on effects of the non-axisymmetric
DDIs. This condition can be experimentally achieved by tuning the amplitude
(Rabi frequency) and the detuning of the $\pi$-field, while keeping the $%
\sigma$-field parameters fixed~\cite{experiment}.

In Sec.~III, we first fix the effective scattering length $a_s$
and vary the particle number $N$ and the
relative strength $\epsilon_{dd}^{(2)}$ of the non-axisymmetric DDI.
In Sec.~IV, we consider a second idealized parameter cut in which
$a_{d2}$ and $N$ are fixed while $a_s$ is varied.

\section{Ground-State (GS) Properties}

\subsection{\NoCaseChange{Formation of QDs}}

\begin{figure}[tbp]
\centering
\includegraphics[width=0.48\textwidth, trim=0cm 0cm 0cm 0cm, clip]{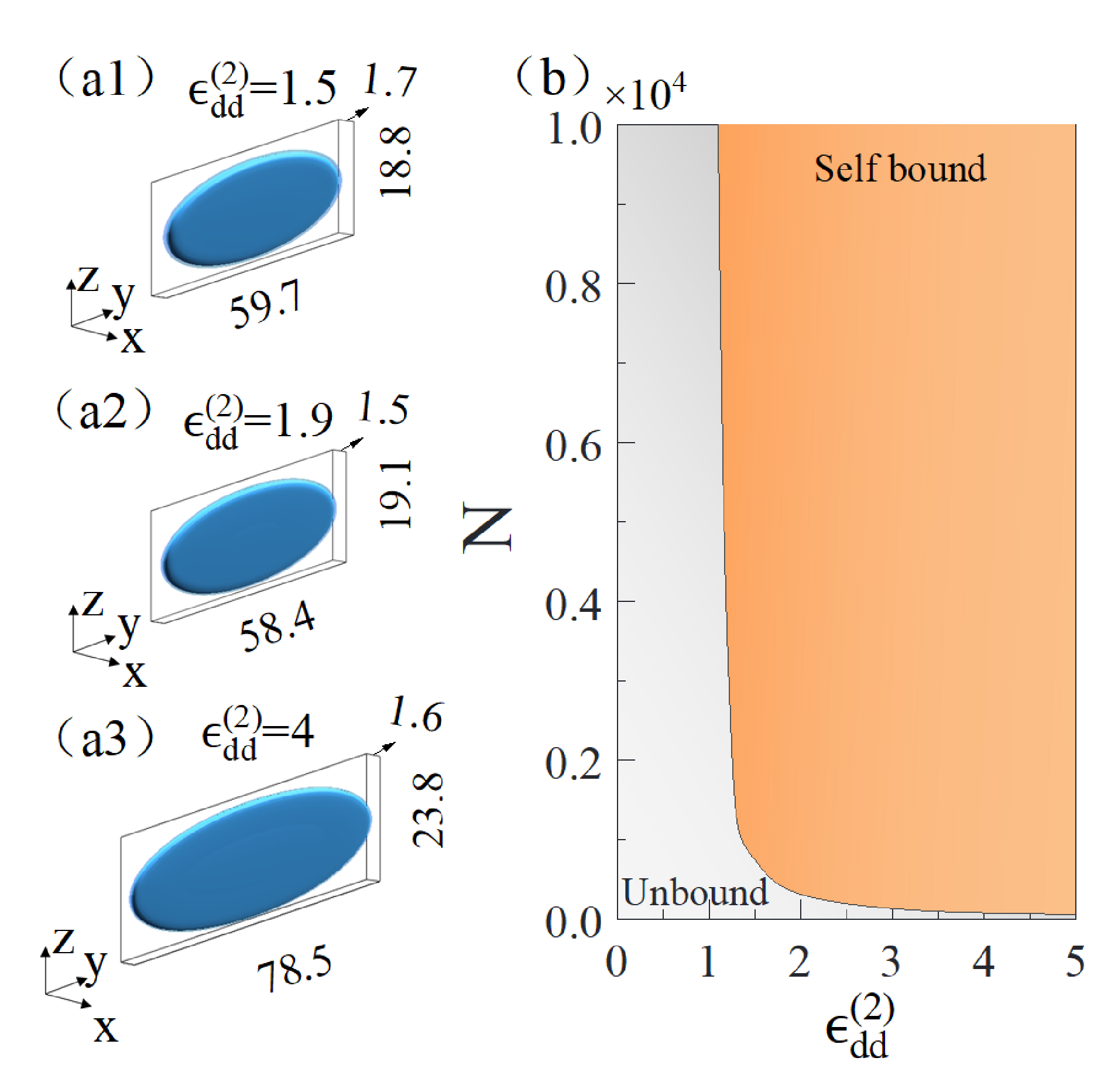}
\caption{ (a1)--(a3) Isodensity profiles of self-bound QDs for $N=10000$
with $\protect\epsilon _{dd}^{(2)}=1.5$, $1.9$, and $4$, respectively. (b)
The existence region of ground-state self-bound QDs for $\protect\epsilon %
_{dd}^{(0)}=0$ and $a_{s}=2100\,a_{0}$. The boundary (black line) separates
unbound (grey) and self-bound (orange) regions. }
\label{fig:Fin2}
\end{figure}

In this section, we adopt the representative value $a_s=2100\,a_0$,
 which is taken from the experimental data in Ref.~\cite{experiment},
  where $a_{0}$ is the Bohr radius, the corresponding value of the coefficient
  in Eq.~(\ref{stationary-GPE}) being $g_s=4\pi a_s/l_{0}=0.7$. The GS-QD state was
produced by Eq. by means of the imaginary-time
method~\cite{ITP}. Stability of the QDs was tested by direct simulations of
their perturbed evolution (not shown here in detail), confirming that all
QDs are stable, as might be expected. Typical examples of the QDs for
different values of $\epsilon_{dd}^{(2)}$ are shown in Fig.~\ref{fig:Fin2}%
(a1)--\ref{fig:Fin2}(a3), with $N=10000$ and $\epsilon_{dd}^{(2)}=1.5$, $1.9$%
, and $4$, respectively. As $\epsilon_{dd}^{(2)}$ increases, the QD becomes
denser at intermediate interaction strengths and gradually develops a more
anisotropic shape at larger $\epsilon_{dd}^{(2)}$.

Fig.~\ref{fig:Fin2}(b) shows the existence region of self-bound QDs for $%
\epsilon _{dd}^{(0)}=0$ and $a_{s}=2100\,a_{0}$. Stable self-bound states
exist only in the orange region above the boundary (black line), whereas the
grey region corresponds to unbound states where no bound-state solution is
found. Self-bound states are not supported for sufficiently small particle
numbers $N$ or small values of $\epsilon _{dd}^{(2)}$. In particular, the
limit of $\epsilon _{dd}^{(2)}=0$ does not support self-bound states because
the effective mean-field interaction becomes purely repulsive ($g_{s}>0$).
At $N=1000$, the threshold found in Fig.~\ref{fig:Fin2}(b) is consistent
with the result reported by Baillie \cite{Baillie}. Notably, $E<0$ in Fig.~%
\ref{fig:Fin2}(b) confirms the stability of the QDs against decay.

As shown in Fig. \ref{fig:Fin2}, under the condition
$\epsilon^{(0)}_{dd}\equiv0$ and $\epsilon^{(2)}_{dd}\neq0$,
the droplets exhibit a highly anisotropic, elliptical pancake-like
density profile, characterized by a narrow width along the $x$ direction
and an extended, nearly flat-top density distribution in $(y,z)$ plane.
In Ref. \cite{Baillie}, an anisotropic Gaussian variational ans\"{a}tz
was employed to characterize the GS properties of microwave-dressed
dipolar quantum droplets in the general case, where both $\epsilon^{(0)}_{dd}$
nd $\epsilon^{(2)}_{dd}$ are nonzero. While this ans\"{a}tz provides a useful
description of the overall anisotropic structure, it does not fully capture
the pronounced flat-top feature of the density profile in the purely
non-axisymmetric regime considered here. To better account for this
characteristic droplet shape, we introduce an anisotropic
Gaussian--super-Gaussian variational ansatz.
\begin{equation}
\varphi_{\mathrm{VA}}(\mathbf{r})
=Ae^{-\left(\frac{x}{w_x}\right)^2}
e^{-\left[
\left(\frac{y}{w_y}\right)^{\alpha_y}
+\left(\frac{z}{w_z}\right)^{\alpha_z}
\right]},
\label{eq:va_ansatz}
\end{equation}

where $w_x$, $w_y$, and $w_z$ are the variational widths ($w_{x,y,z}>0$),
 while $\alpha_y$ and $\alpha_z$ control the flatness of the density
  profile along the $y$ and $z$ directions, respectively, with
  $\alpha_{y,z}\geq2$. The amplitude $A$ is determined by the
  particle number $N$ and the variational parameters.

Substituting Eq.~\eqref{eq:va_ansatz} into the energy functional
 in Eq.~\eqref{E}, the total energy is expressed as a function of
 variational parameters $(w_x,w_y,w_z,\alpha_y,\alpha_z)$. The kinetic,
 contact-interaction, and LHY contributions are evaluated analytically
 from the corresponding integrals of the trial state. The dipolar
 contribution is evaluated numerically using FFT convolution with the
 same spherical cutoff as in the imaginary-time calculations. The
 variational parameters are then determined by numerically minimizing
 the total energy using the { \texttt{fminsearch} routine~\cite{MATLAB_fminsearch}}.

\begin{figure}[tbp]
\centering
\includegraphics[width=0.42\textwidth, trim=0cm 0cm 0cm 0cm, clip]{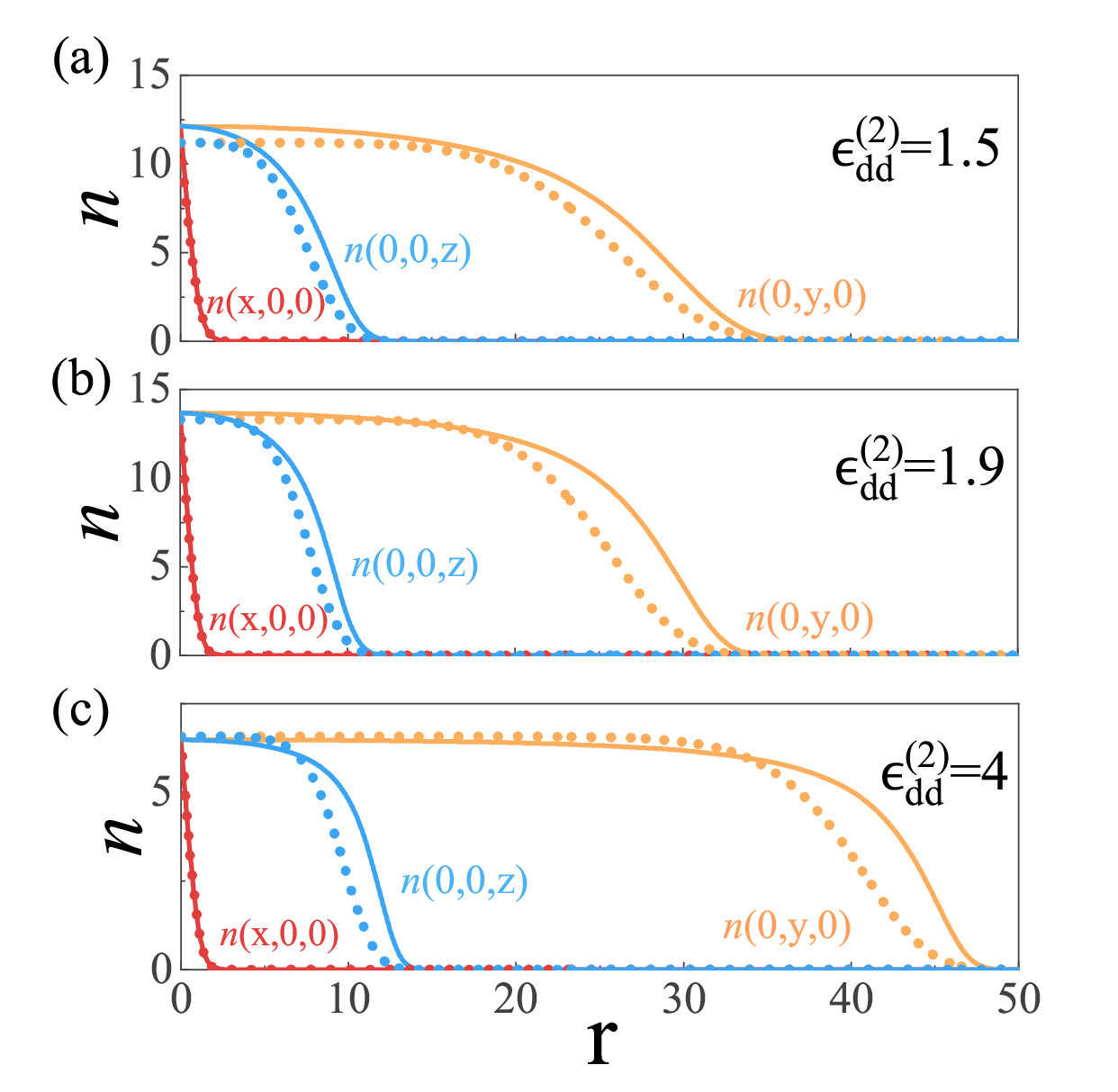}
\caption{Density profiles along the $x$ (red), $y$ (yellow), and $z$ (blue)
directions for the self-bound QDs shown in Fig.~\protect\ref{fig:Fin2}(a).
Solid and dotted curves represent the numerical and the
VA results, respectively. Panels (a)--(c) correspond to
$\epsilon_{dd}^{(2)}=1.5$, $1.9$, and $4$, respectively, with $N=10000$.
The optimized variational parameters $(w_x,w_y,w_z,\alpha_y,\alpha_z)$ are
$(0.97,30.49,9.38,5.98,4.46)$,
$(0.82,29,9.19,7.17,5.3)$, and
$(0.82,43.63,11.09,12,7.07)$ for panels (a)--(c), respectively.}
\label{fig:Fin3}
\end{figure}

In Fig.~\ref{fig:Fin3}, the dashed lines represent the variational profiles.
Overall, the variational results agree well with the imaginary-time profiles
over the main density region, with a slight overestimation of the low-density
tails. The variational results yield $\alpha_y,\alpha_z$ well above 2,
 confirming the pronounced flat-top character of the density profiles
 in the $(y,z)$ plane and supporting the validity of the Gaussian-super-Gaussian
  ansatz for describing the droplet structure.

For all considered interaction strengths, the QDs exhibit pronounced
 directional dependence, with the largest spatial extent along the $y$
 direction. As $\epsilon _{dd}^{(2)}$ increases, the density profiles
 first broaden and then become more localized, indicating a
nonmonotonic evolution of the QD size.
 This observation motivates a quantitative
 analysis of the QD properties. We use chemical potential $\mu$ and total
 energy $E$ to characterize the cohesion of the self-bound state, and the
 effective volume $V$ and peak density $n_p$ to quantify its spatial localization
 and density compression. We further compare these numerical
 results with the variational approximation (VA) introduced above to assess
 the capability of the Gaussian-super-Gaussian ans\"{a}tz in describing the QD properties.
Firstly, we examine how these quantities depend on
 $\epsilon _{dd}^{(2)}$ and $N$.

\begin{figure}[tbp]
\centering
\includegraphics[width=0.48\textwidth, trim=0cm 0cm 0cm 0cm, clip]{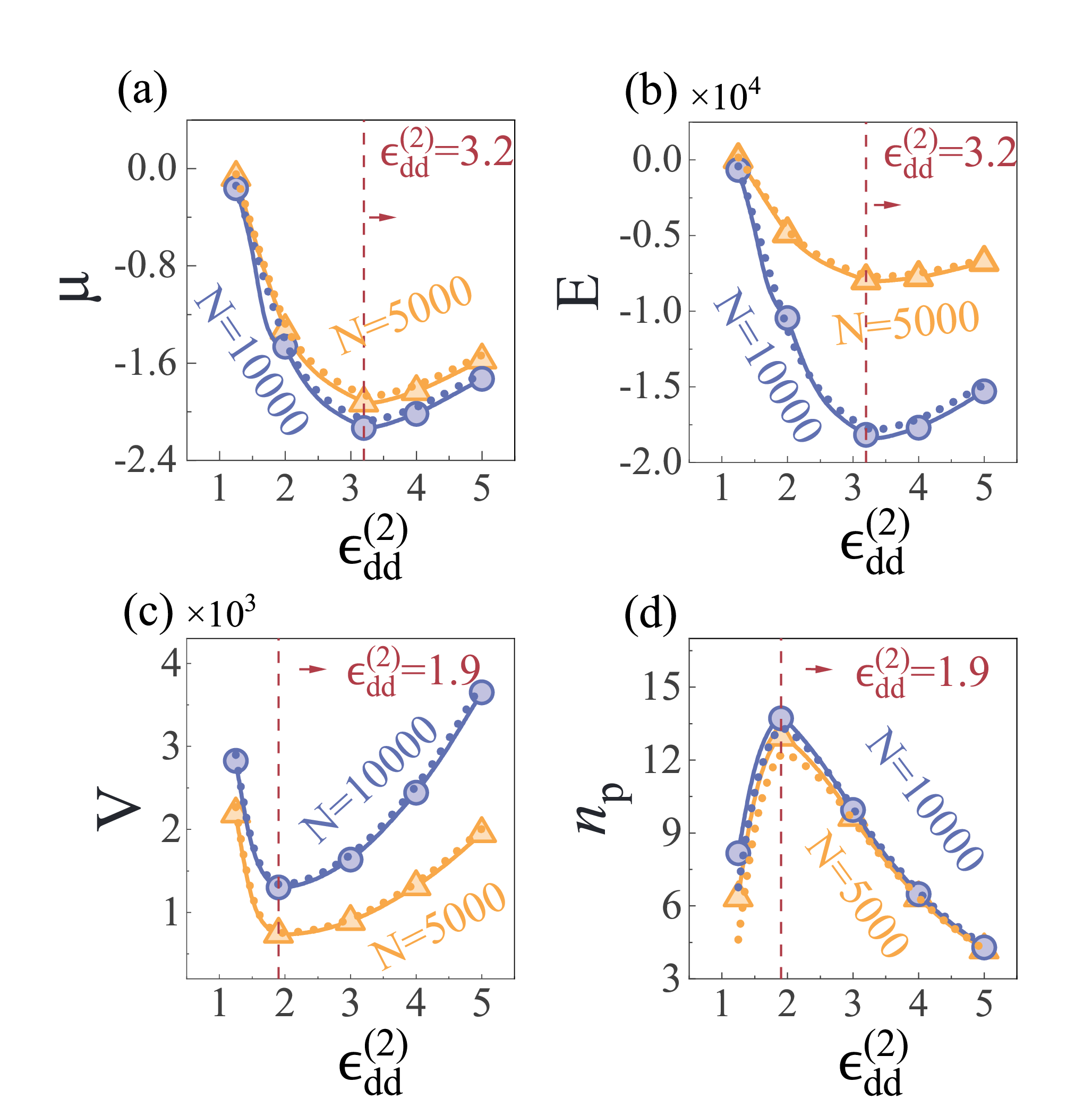}
\caption{QD properties as functions of the non-axisymmetric DDI strength, $%
\protect\epsilon _{dd}^{(2)}$, for $N=5000$ (yellow) and $N=10000$ (purple),
respectively. (a) Chemical potential $\protect\mu $. (b) Total energy $E$.,
see Eq. (\protect\ref{E}). (c) Effective volume $V$, see Eq. (\protect\ref{V}%
). (d) Peak density $n_{p}$. Solid and dotted lines
 denote the numerical and VA results, respectively. }
\label{fig:Fin4}
\end{figure}

As shown in Figs.~\ref{fig:Fin4}(a) and (b), the solid
and dotted curves represent the numerical and VA results, respectively. For fixed
particle numbers $N=5000$ and $N=10000$, both the chemical potential $\mu $
and the total energy $E$ exhibit a nonmonotonic trend with increasing
$\epsilon _{dd}^{(2)}$.

The two quantities first decrease and reach minima
near $\epsilon _{dd}^{(2)}\approx 3.2$, indicating the strongest self-binding
within the sampled range.
For larger $\epsilon _{dd}^{(2)}$, both quantities
gradually increase, signaling weakened effective binding. The minima occur at
nearly the same value for the two particle numbers shown here, suggesting only
a weak $N$ dependence in this parameter window; a denser scan would be needed
to establish the precise dependence of the minimum position on $N$.

Besides the energetic properties, it is also instructive to examine how the
density and size of the QD evolve. As shown in Fig.~\ref{fig:Fin4}(c), the
effective volume $V$ first increases with $\epsilon_{dd}^{(2)}$, reaches a
maximum near $\epsilon_{dd}^{(2)}\approx1.9$, and then decreases at larger
anisotropy. A correlated behavior is observed in the peak density $n_p$
shown in Fig.~\ref{fig:Fin4}(d), which first decreases and subsequently
increases at stronger anisotropy, reflecting initial expansion of the QD
followed by compression at large $\epsilon_{dd}^{(2)}$.

\begin{figure}[tbp]
\centering
\includegraphics[width=0.48\textwidth, trim=0cm 0cm 0cm 0cm, clip]{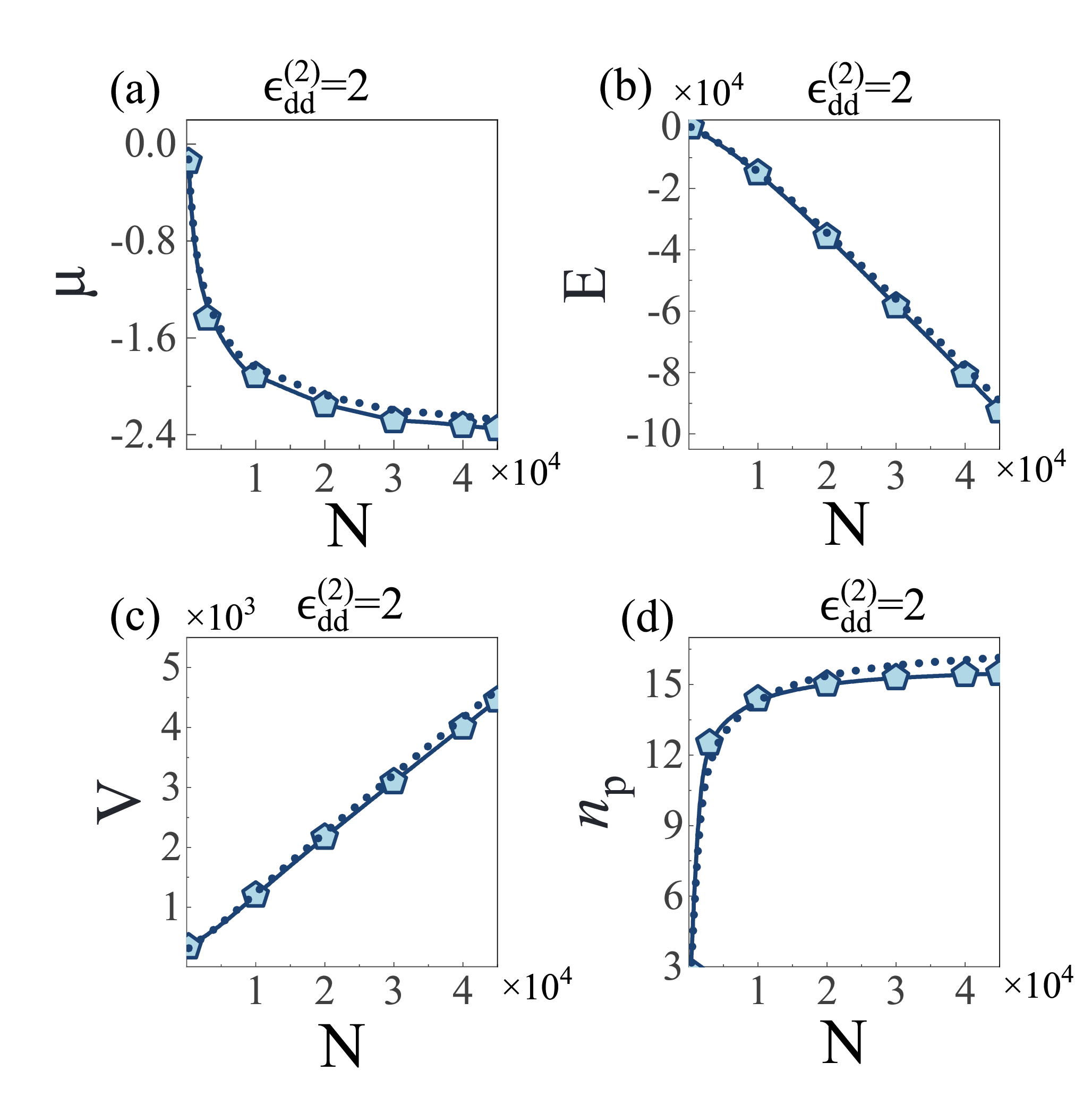}
\caption{QD properties as functions of the particle number, $N$, for $%
\protect\epsilon _{dd}^{(0)}=0$ and $\protect\epsilon _{dd}^{(2)}=2$. (a)
Chemical potential $\protect\mu $. (b) Total energy $E$, see Eq. (\protect
\ref{E}). (c) Effective volume $V$, see Eq. (\protect\ref{V}). (d) Peak
density $n_{p}$. Solid and dotted lines denote the numerical and VA results, respectively. }
\label{fig:Fin5}
\end{figure}

To further characterize the basic properties of self-bound QDs, we next
select a representative $\epsilon_{dd}^{(2)}=2$ and investigate their
dependence on particle number $N$. As shown in Figs.~\ref{fig:Fin5}(a) and %
\ref{fig:Fin5}(b), the chemical potential $\mu$ decreases while the total
energy $E$ becomes increasingly negative with increasing $N$, reflecting
enhanced self-binding. The variation is rapid at small particle number and
gradually slows at around $N=10000$, approaching an approximately linear
dependence within the explored range.

Importantly, the condition $d\mu /dN<0$ is satisfied throughout the explored
parameter range, consistent with the Vakhitov-Kolokolov (VK) stability
criterion for self-bound states~\cite{VK}. This behavior suggests a
crossover from a dilute regime at small particle numbers to a denser and
more weakly compressible QD state at larger $N$.

At the same representative $\epsilon _{dd}^{(2)}=2$, we next examine how the
density and size properties evolve with $N$. As shown in Fig.~\ref{fig:Fin5}%
(c), the peak density $n_{p}$ with $N$. Meanwhile, the effective volume $V$
in Fig.~\ref{fig:Fin5}(d) increases rapidly at small particle number and
then grows more slowly as $N$ keeps growing, suggesting that additional
particles contribute to expansion of the QD, rather than increase of its
density.  This behavior is consistent with the flat-top
regime of self-bound quantum droplets~\cite{Petrov2015}.

Having discussed the basic properties of the QD, we now focus on its
geometric anisotropy induced by the non-axisymmetric dipole-dipole
interaction. To characterize the anisotropic deformation, we define the
anisotropy ratios

\begin{equation}
\eta_{zy}=W_z/W_y,
\end{equation}
\begin{equation}
\eta_{xy}=W_x/W_y,
\end{equation}

where the effective lengths along the three spatial directions are defined
as
\begin{equation}
W_x = \frac{\left(\int |\varphi(x,0,0)|^2 \, dx\right)^2}{\int
|\varphi(x,0,0)|^4 \, dx},
\end{equation}
\begin{equation}
W_y = \frac{\left(\int |\varphi(0,y,0)|^2 \, dy\right)^2}{\int
|\varphi(0,y,0)|^4 \, dy},
\end{equation}
\begin{equation}
W_z = \frac{\left(\int |\varphi(0,0,z)|^2 \, dz\right)^2}{\int
|\varphi(0,0,z)|^4 \, dz}.
\end{equation}

This definition is convenient for the present flat-top droplets:
in the Thomas-Fermi limit, where the central density is approximately
constant within the half-width \(R_i^{\rm TF}\) and vanishes outside it, the
above definition gives \(W_i=2R_i^{\rm TF}\)~\cite{Zhao2026}. Thus, \(W_i/W_j\) directly
measures the corresponding Thomas-Fermi aspect ratio.

Here, $W_y$ is chosen as the reference length since the QD generally has the
largest spatial extent along the $y$ direction. Therefore, smaller values of
$\eta_{zy}$ and $\eta_{xy}$ indicate stronger anisotropy, corresponding to a
more elongated density distribution along $y$.

\begin{figure}[]
\centering
\includegraphics[width=0.5\textwidth, trim=0cm 0cm 0cm 0cm, clip]{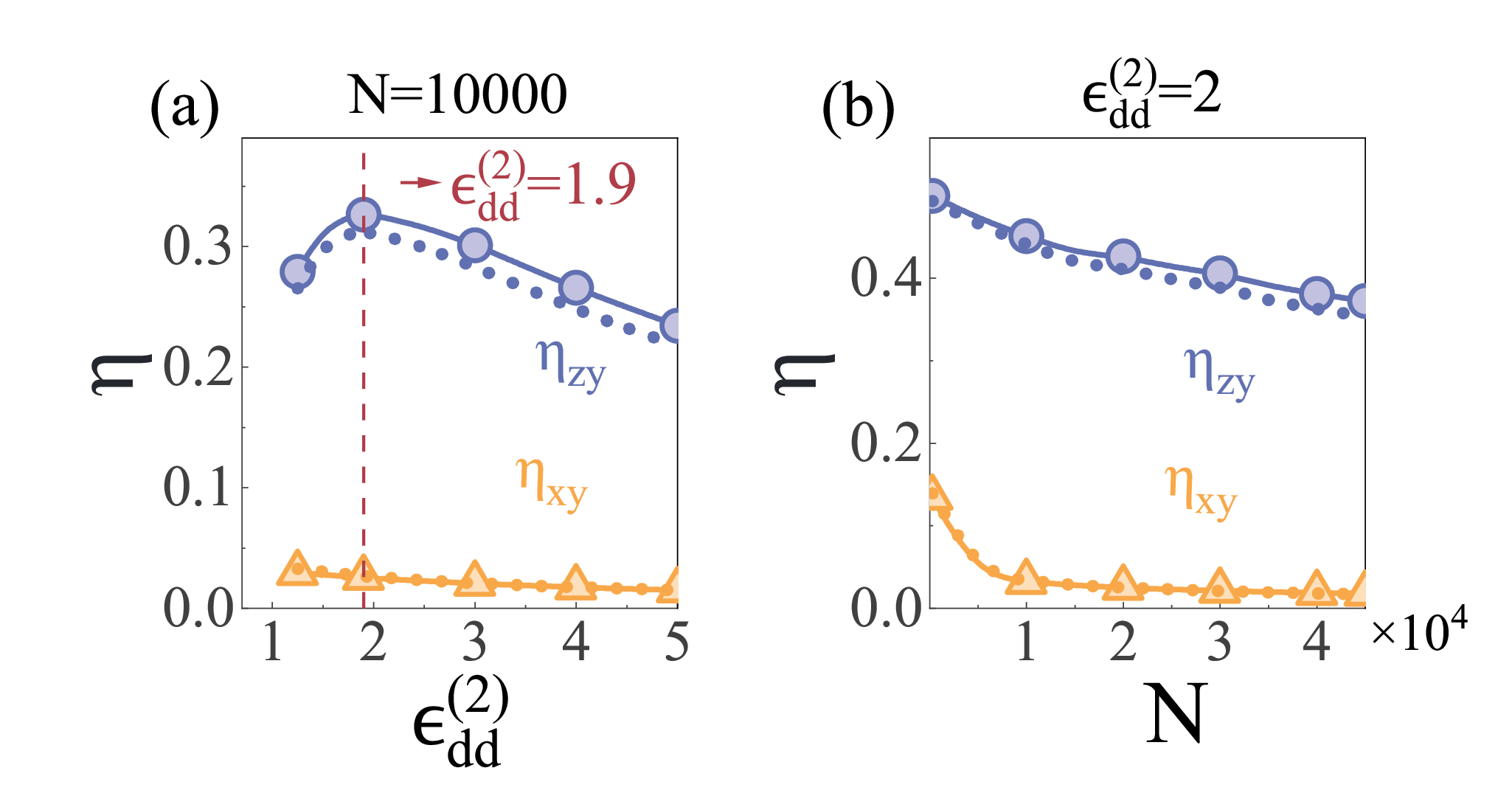}
\caption{ (a) Anisotropy ratios $\protect\eta_{zy}$ (purple) and $\protect%
\eta_{xy}$ (yellow) as a function of $\protect\epsilon_{dd}^{(2)}$ with $%
N=10000$. (b) Anisotropy ratios $\protect\eta_{zy}$ (purple) and $\protect%
\eta_{xy}$ (yellow) as a function of $N$ with $\protect\epsilon_{dd}^{(2)}=2$%
. Solid and dotted lines denote the numerical and VA results, respectively. }
\label{fig:Fin6}
\end{figure}

Fig.~\ref{fig:Fin6}(a) and Fig.~\ref{fig:Fin6}(b) show the anisotropy ratios
as functions of $\epsilon_{dd}^{(2)}$ and $N$, respectively.
Since increasing $\epsilon_{dd}^{(2)}$ strengthens the anisotropic dipolar
interaction, while increasing $N$ enhances the overall nonlinear interaction
energy, the QD is expected to become increasingly elongated along the $y$
direction. As a result, both $\eta_{zy}$ and $\eta_{xy}$ are generally
expected to decrease with increasing $\epsilon_{dd}^{(2)}$ or $N$,
reflecting enhanced geometric anisotropy. Overall, the numerical results
follow this tendency.

However, a slight deviation from this trend appears near $%
\epsilon_{dd}^{(2)}=1.9$, where $\eta _{zy}$ exhibits a small local maximum,
corresponding to a relatively thicker density profile along the $z$
direction. This behavior may originate from the competition between the
anisotropic DDI and the stabilizing effects of the contact and LHY
interactions, which can temporarily weaken the effective compression along
the $z$ direction.

Overall, the VA results agree well with the numerical results over the
explored parameter ranges, demonstrating that the present Gaussian-super-Gaussian
ansatz captures well the main properties of the QDs.

\subsection{\NoCaseChange{Contact Interaction Tuning at Fixed $a_{\mathrm{d2}}$}}

To further clarify the role of the contact interaction
in the formation of self-bound quantum droplets, we consider the dependence
of the droplet properties on the $s$-wave scattering length $a_s$ while
keeping the dipolar length $a_{\mathrm{d2}}$ fixed. Although the independent
experimental control of these two interaction parameters may require appropriate
 tuning of the microwave dressing fields, a reasonable schemes for their
  independent tunability have been discussed in the Ref. \cite{Singh2026}.
   Here, we therefore treat $a_s$ and $a_{\mathrm{d2}}$ as independently
   controllable parameters and focus on the theoretical role of the contact
   interaction by varying $a_s$ at a fixed $a_{\mathrm{d2}}$. This provides
   a direct way to isolate how the short-range contact interaction affects
   the formation and properties of the self-bound droplets.

Specifically, we fix the non-axisymmetric dipolar length at
$a_{\mathrm{d2}}=4200a_0$ and set $N=2000$, while varying $a_s$. This defines
an idealized fixed-$a_{\mathrm{d2}}$ path within the present eGPE model.
For $\epsilon_{\mathrm{dd}}^{(0)}=0$, substituting
$G_s=4\pi\hbar^2a_s/m$ and
$\epsilon_{\mathrm{dd}}^{(2)}=a_{\mathrm{d2}}/a_s$ into
Eq.~(\ref{DDI}) gives
\begin{equation}
\begin{aligned}
U_{\mathrm{dd}}(\mathbf R)
&=\frac{3\sqrt{3}G_s\epsilon_{\mathrm{dd}}^{(2)}}{4\pi R^3}
\sin^2\theta\cos 2\phi \\
&=\frac{3\sqrt{3}\hbar^2a_{\mathrm{d2}}}{mR^3}
\sin^2\theta\cos 2\phi.
\end{aligned}
\end{equation}

Thus, the non-axisymmetric mean-field DDI kernel automatically
unchanged along this path, while the contact-interaction coefficient $G_s$
and the LHY coefficient $\gamma_{\mathrm{QF}}$ vary with $a_s$.

\begin{figure}[tbp]
\centering
\includegraphics[width=0.42\textwidth, trim=0cm 0cm 0cm 0cm, clip]{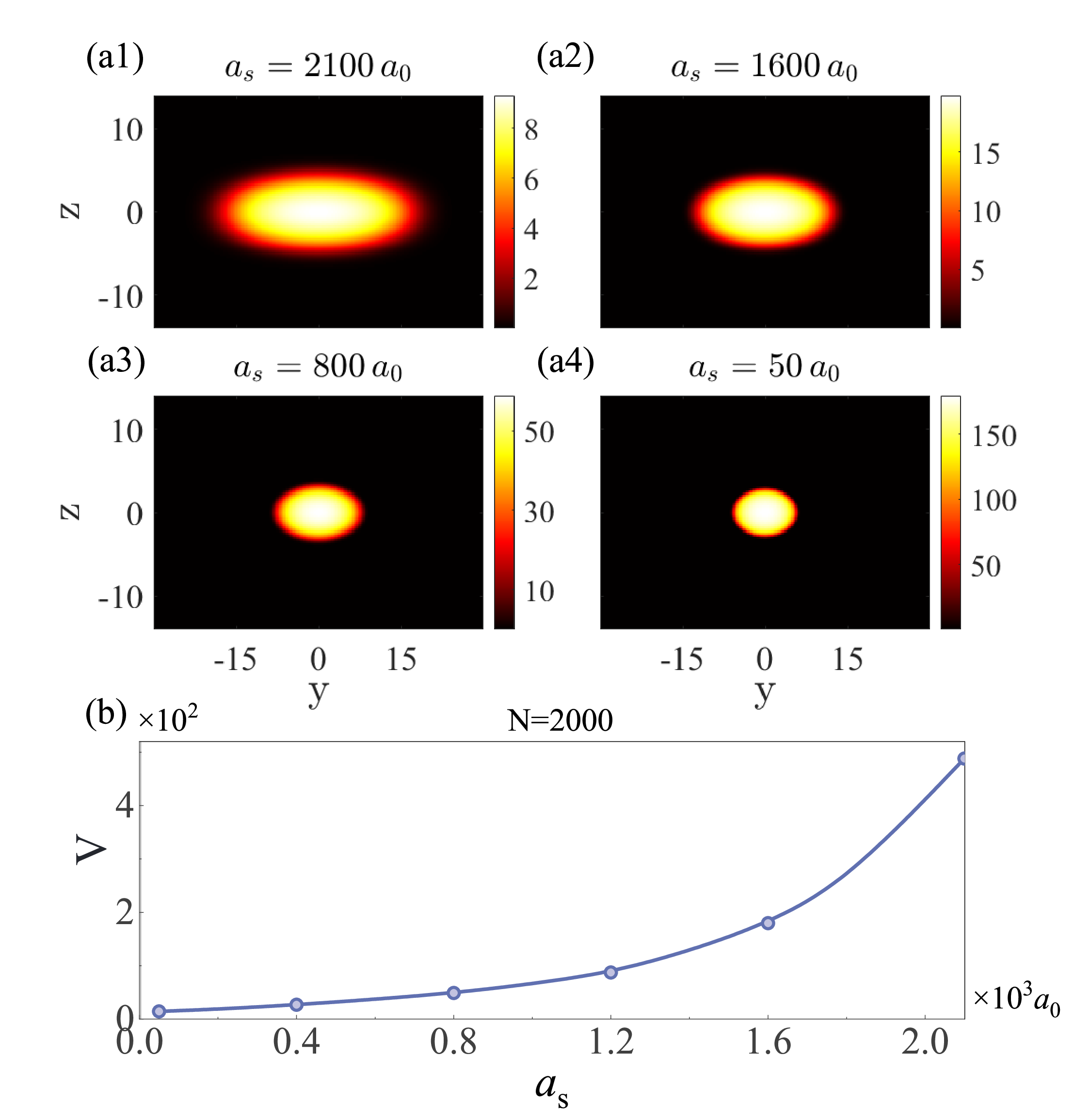}
\caption{ (a1)--(a4) Density distributions in the central $y$-$z$ plane for $%
a_{s}=2100\,a_{0}$, $a_{s}=1600\,a_{0}$, $800\,a_{0}$, and $50\,a_{0}$,
respectively, with fixed $a_{\mathrm{d2}}=4200\,a_{0}$ and $N=2000$. (b)
Effective volume $V$ [see Eq. (\protect\ref{V})] as a function of the $s$%
-wave scattering length $a_{s}$. }
\label{fig:Fin7}
\end{figure}

Figure~\ref{fig:Fin7} summarizes the dependence of the QD properties on
$a_s$. Panels (a1)--(a4) show the density distributions in the central
$x=0$ plane, corresponding to the $(y,z)$ plane, for
$a_s=2100a_0$, $1600a_0$, $800a_0$, and $50a_0$, respectively. Panel (b)
shows the corresponding effective volume $V$ as a function of $a_s$.

Over the parameter range shown, decreasing $a_s$ produces a continuous
compression of the QD, as indicated by the decreasing effective volume and
increasing peak density. These trends indicate increasing compression and
suggest a possible approach to collapse for $a_{s}\rightarrow 0$.
The stationary states obtained by ITP for the parameters shown in Fig.~\ref{fig:Fin7}
remain stable in direct simulations for a sufficiently long time.
Note that, in this figure, as $a_s$ becomes very small, the increasingly
compressed QDs require progressively finer spatial resolution, and the
 numerical results become sensitive to the grid size; we therefore
 terminate the numerical scan the $a_s$ at a small finite value.

\section{\NoCaseChange{DYNAMICS of QDs }}

\subsection{\NoCaseChange{Collision dynamics}}

To investigate the dynamical properties of self-bound QDs, we consider
head-on collisions between two identical QDs with $N=10000$, $%
\epsilon_{dd}^{(2)}=2$, and $a_{s}=2100\textcolor{blue}{\,}a_{0}$. The initial states are
prepared by imprinting opposite momenta onto two spatially separated
stationary QDs.

For a head-on collision along a coordinate axis
$j=x,y,z$, the initial state is written once in the form
\begin{equation}
\begin{gathered}
\psi(\mathbf{r},t=0)=\varphi_1(\mathbf{r}-\mathbf{r}_0)
e^{i\boldsymbol{\zeta}_j\cdot\mathbf{r}}\\
\mathrel{\phantom{\psi(\mathbf{r},t=0)=}}+
\varphi_2(\mathbf{r}+\mathbf{r}_0)
e^{-i\boldsymbol{\zeta}_j\cdot\mathbf{r}},
\end{gathered}
\end{equation}

where $\mathbf{r}=(x,y,z)$, $\mathbf{r}_0=r_{0j}
\hat{\mathbf e}_j$ is the initial displacement of one droplet, and
$\boldsymbol{\zeta}_j=\zeta_j\hat{\mathbf e}_j$ is the corresponding
kick vector. The parameters are $r_{0x}=12$ for the $x$ collision,
$r_{0y}=128$ for the $y$ collision, and $r_{0z}=64$ for the $z$ collision.
The initial displacements are chosen to avoid significant density overlap at
$t=0$.

\begin{figure}[tbp]
\centering
\includegraphics[width=0.44\textwidth, trim=0cm 0cm 0cm 0cm, clip]{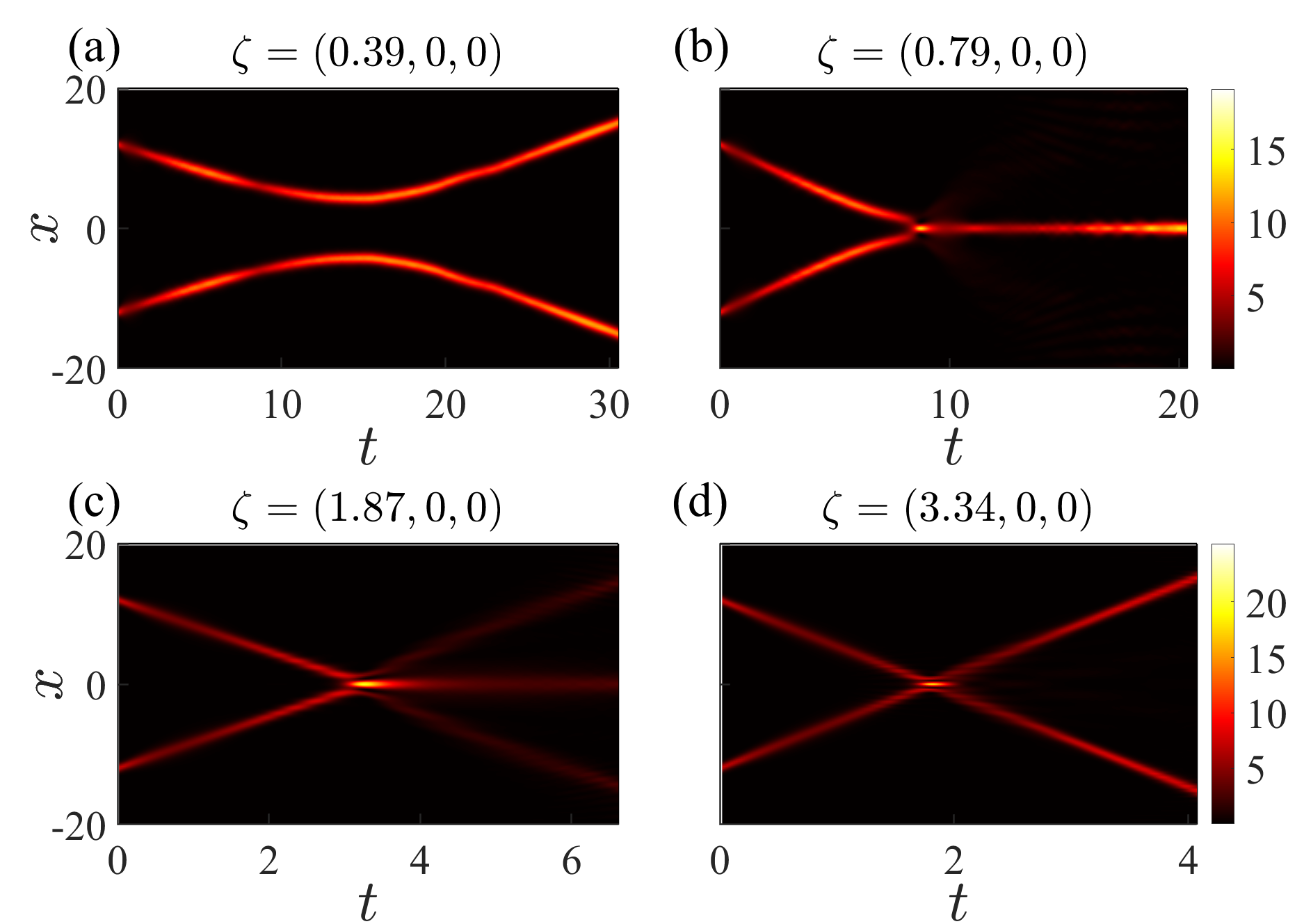}
\caption{Time evolution of the central density profiles during head-on
collisions along the $x$ direction of two identical
self-bound QDs with $N=10000$, $\protect\epsilon _{dd}^{(2)}=2$, $%
a_{s}=2100\,a_{0}$, and $r_{0x}=12$.
The kick strength is (a) $\zeta_x=0.39$, (b) $\zeta_x=0.79$,
(c) $\zeta_x=1.87$,
and (d) $\zeta_x=3.34$.}
\label{fig:Fin8}
\end{figure}

For collisions along the $x$ direction, the density cross
sections in the $(x,t)$-plane, shown in Figs.~\ref%
{fig:Fin8}(a)--\ref{fig:Fin8}(d), clearly indicate that the collision
dynamics strongly depend on the kick strength, $\zeta_x$. For weak kicks with
$\zeta_x=0.39$, shown in Fig.~\ref{fig:Fin8}(a), the QDs
approach each other and subsequently reverse their motion before substantial
density overlap occurs, producing a quasi-elastic rebound with only weak
deformation. For stronger kicks with $\zeta_x=0.79$, shown
in Fig.~\ref{fig:Fin8}(b), the QDs merge upon collision and subsequently
form a single self-bound QD.

For a still larger kick with $\zeta_x=1.87$, shown in
Fig.~\ref{fig:Fin8}(c), the collision still produces a merged central droplet,
but two weak emitted branches also appear and propagate away from the collision
region. At the largest kick considered here, $\zeta_x=3.34$, shown in
Fig.~\ref{fig:Fin8}(d), the central merged component is almost suppressed after
the collision, and the final density distribution is dominated by two outgoing
emission branches.

\begin{figure}[tbp]
\centering
\includegraphics[width=0.44\textwidth, trim=0cm 0cm 0cm 0cm, clip]{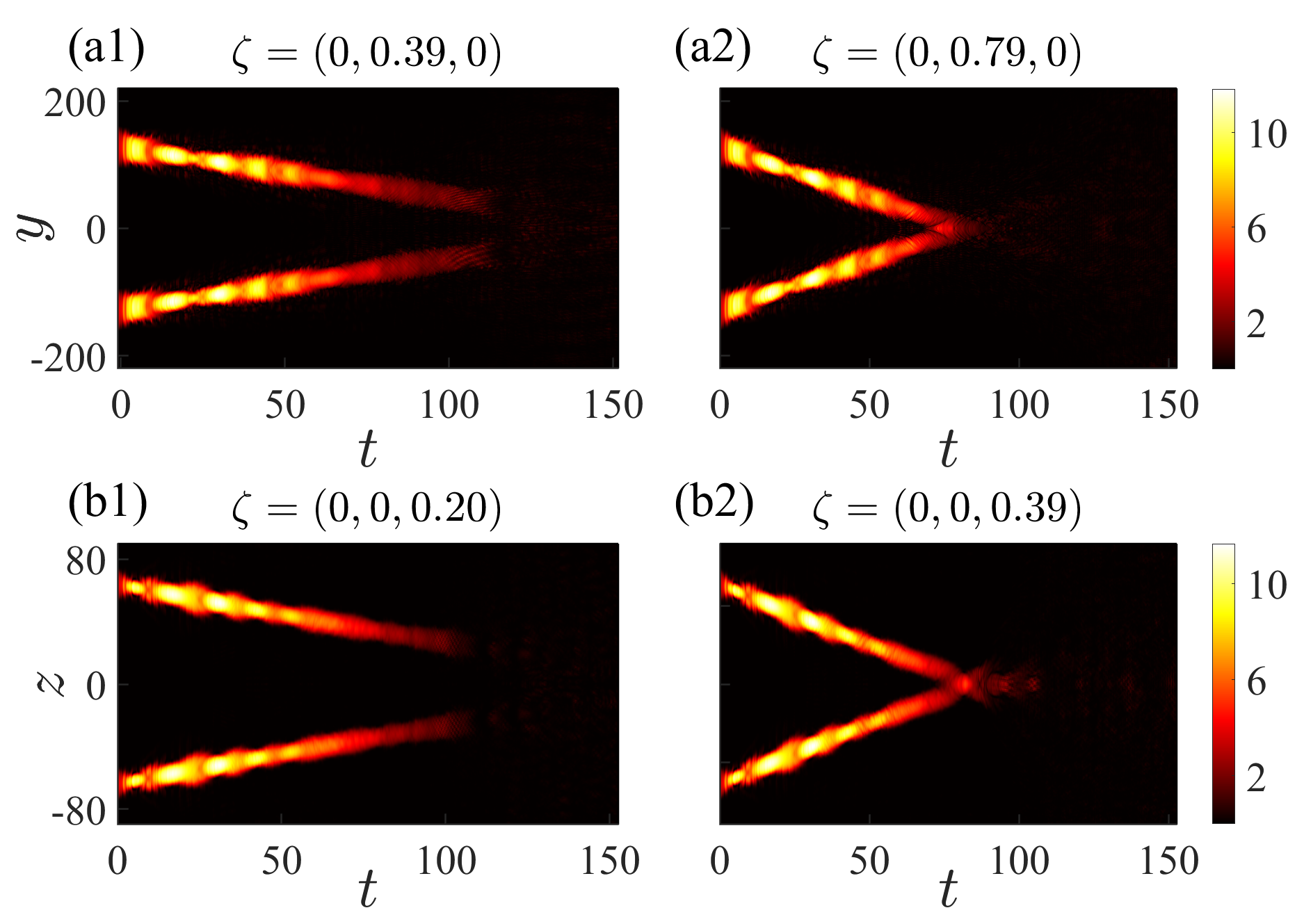}
\caption{Time evolution of the central density profiles during head-on
collisions of two identical self-bound QDs with $N=10000$, $\protect\epsilon %
_{dd}^{(2)}=2$, and $a_{s}=2100\textcolor{blue}{\,}a_{0}$ for different collision directions and
kick strengths. (a1)--(a2) Collisions along the $y$ direction for $\protect\zeta_y=0.39$
and $\protect\zeta_y=0.79$, respectively, with $r_{0y}=128$.
 (b1)--(b2) Collisions along the $z$ direction for
 $\protect\zeta_z=0.20$ and $\protect\zeta_z=
0.39$, respectively, with $r_{0z}=64$. }
\label{fig:Fin9}
\end{figure}

The $(y,t)$- and $(z,t)$ -plane density cross sections in
Figs.~\ref{fig:Fin9}(a1)--\ref{fig:Fin9}(a2) and~\ref{fig:Fin9}(b1)--\ref{fig:Fin9}(b2) show that the
collisions are likewise
strongly inelastic  in both transverse directions. For the weaker kicks
$\zeta_y=0.39$ in the $y$ direction and $\zeta_z=0.20$ in the $z$ direction,
shown in Figs.~\ref{fig:Fin9}(a1) and~\ref{fig:Fin9}(b1), respectively,
the QDs deform and fragment before substantial overlap. For the stronger
kicks $\zeta_y=0.79$ and $\zeta_z=0.39$, shown in Figs.~\ref{fig:Fin9}(a2) and
\ref{fig:Fin9}(b2), respectively, the QDs first overlap and then break into
fragments. No long-lived self-bound remnant is observed in either direction.

\subsection{Rotational dynamics}

We investigate the rotational response of a single self-bound QD to a
time-dependent anisotropic DDI. We obtain a stationary QD
 for $N=5000$, $\epsilon_{dd}^{(0)}=0$, $\epsilon_{dd}^{(2)}=2$, and $a_s=2100a_0$.
The stationary state serves as the initial state for real-time evolution
 with a continuously rotating anisotropic DDI about one of the three Cartesian
  axes.

The rotation angle evolves as $\beta(t)=\Omega t$, where $\Omega$ is the angular velocity.
The time-dependent orientation of the anisotropic DDI is implemented by transforming the
Fourier-space momentum vector $\mathbf{k}$ according to

\begin{equation}
\mathbf{k}_{\beta}^{\{x,y,z\}}
=\mathsf{T}_{\{x,y,z\}}\!\left[\beta(t)\right]\mathbf{k},
\label{eq:rotated_k}
\end{equation}

where $\mathsf{T}_{\{x,y,z\}}$ denotes the rotation matrix about the corresponding
Cartesian axis, and $k_{j,\beta}^{\{x,y,z\}}$ is the $j$th component of
$\mathbf{k}_{\beta}^{\{x,y,z\}}$ with $j=x,y,z$.

The corresponding transformed components are
\begin{align}
\mathbf{k}_{\beta}^{x}
&=\left(k_{\beta,x}^{x},\,k_{\beta,y}^{x},\,k_{\beta,z}^{x}\right)
\nonumber\\
&=\bigl(k_x,\,
k_y\cos\beta-k_z\sin\beta,\,
k_y\sin\beta+k_z\cos\beta\bigr),
\label{eq:rotated_components_x}\\[4pt]
\mathbf{k}_{\beta}^{y}
&=\left(k_{\beta,x}^{y},\,k_{\beta,y}^{y},\,k_{\beta,z}^{y}\right)
\nonumber\\
&=\bigl(k_x\cos\beta-k_z\sin\beta,\,
k_y,\,
k_x\sin\beta+k_z\cos\beta\bigr),
\label{eq:rotated_components_y}\\[4pt]
\mathbf{k}_{\beta}^{z}
&=\left(k_{\beta,x}^{z},\,k_{\beta,y}^{z},\,k_{\beta,z}^{z}\right)
\nonumber\\
&=\bigl(k_x\cos\beta-k_y\sin\beta,\,
k_x\sin\beta+k_y\cos\beta,\,
k_z\bigr).
\label{eq:rotated_components_z}
\end{align}

Substituting the transformed momentum components into Eq.~(\ref{eq:Fourier_DDI}) gives
\begin{equation}
\begin{aligned}
\tilde{U}_{\mathrm{dd}}(\mathbf{k},t)
=G_s\biggl[
&\epsilon_{dd}^{(0)}
\left(\frac{3\left(k_{\beta,z}^{\{x,y,z\}}\right)^2}{k^2}-1\right)\\
&-\sqrt{3}\epsilon_{dd}^{(2)}
\frac{\left(k_{\beta,x}^{\{x,y,z\}}\right)^2
-\left(k_{\beta,y}^{\{x,y,z\}}\right)^2}{k^2}
\biggr],
\end{aligned}
\label{eq:rot_udd}
\end{equation}

The wave function is propagated using the same split-step Fourier method
as in the collision simulations.The angular velocities are chosen as
$\Omega_x=\Omega_z=2.5\pi\times10^{-3}$ and $\Omega_y=1.25\pi\times10^{-2}.$
To reduce boundary effects, the numerical box is enlarged in the $yz$,
$xz$, and $xy$ planes for rotations about the $x$, $y$, and $z$ axes,
respectively.

\begin{figure}[tbp]
\centering
\includegraphics[width=0.50\textwidth, trim=0cm 0cm 0cm 0cm, clip]{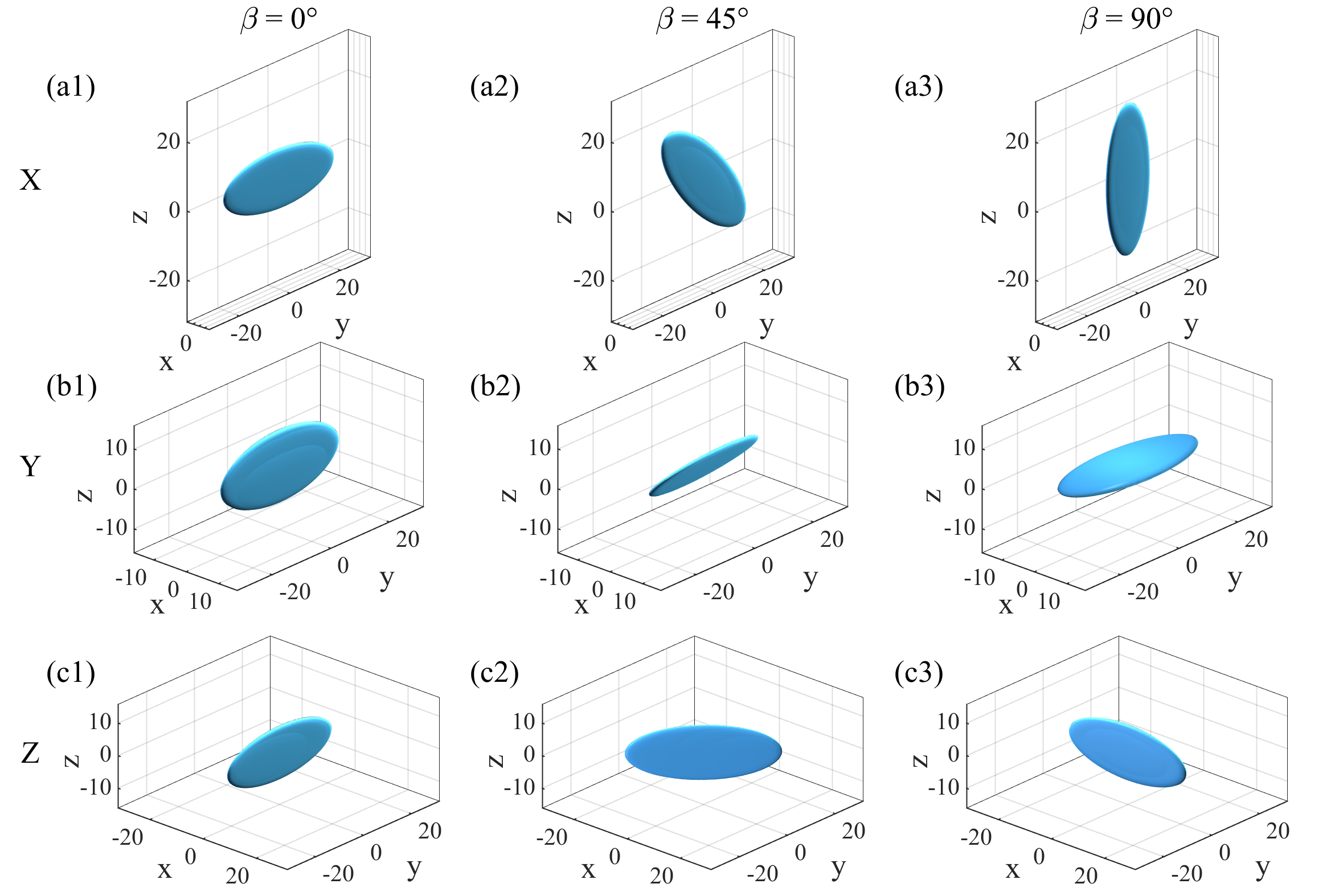}
\caption{Columns show the rotation angles $\beta=0^\circ$,
 $45^\circ$, and $90^\circ$. The plotted surface is the density isosurface
 $n=0.1n_{\mathrm{max}}$. The initial state is the imaginary-time ground state
  for $N=5000$, $\protect\epsilon_{dd}^{(0)}=0$, $\protect\epsilon_{dd}^{(2)}=2$,
  and $a_{s}=2100a_{0}$. Rows (a1)--(a3), (b1)--(b3), and (c1)--(c3) correspond
  to rotations about the $x$, $y$, and $z$ axes, respectively. }
\label{fig:Fin10}
\end{figure}

The resulting density evolution is shown in Fig.~\ref{fig:Fin10}.
The three rows correspond to rotations about the $x$, $y$, and $z$ axes,
 respectively, while the columns show $\beta=0^\circ$, $45^\circ$,
 and $90^\circ$. The density isosurface is plotted at $n=0.1n_{\mathrm{max}}$.
 As $\beta$ increases, the changing orientation of the anisotropic interaction
 induces a corresponding deformation of the self-bound QD, which remains
 localized throughout the evolution.

\section{Conclusion}

We have investigated self-bound quantum droplets (QDs) supported by
non-axisymmetric microwave-dressed dipole-dipole interactions (DDIs) in a
 polar-molecule Bose-Einstein condensate by solving the extended
 Gross-Pitaevskii equation (eGPE) including the Lee-Huang-Yang (LHY)
 correction. We find that stable self-bound QDs can exist even when
 the conventional axially symmetric DDI is completely suppressed, and
 identify their existence region in the $(N,\epsilon_{dd}^{(2)})$ parameter
 plane. To characterize their strongly anisotropic and flattened density
  profiles, we introduce an anisotropic Gaussian-super-Gaussian variational
  ansatz. The optimized variational parameters accurately capture the
  characteristic widths and flat-top features of the QDs, with only small
  deviations in the low-density tails. The variational density profiles are
  in good overall agreement with the numerical eGPE results.

We further characterize the ground-state properties of the QDs through
the chemical potential $\mu$, total energy $E$, effective volume $V$,
peak density $n_p$, and geometric anisotropy as functions of the particle
number $N$ and the non-axisymmetric dipolar interaction strength
$\epsilon_{dd}^{(2)}$. The results reveal pronounced anisotropic
deformation, nonmonotonic dependences of the droplet properties on
$\epsilon_{dd}^{(2)}$, and increasingly strong self-binding with
increasing $N$. The corresponding variational predictions show
good agreement with the numerical simulations, demonstrating that
the Gaussian-super-Gaussian ansatz provides a useful description
of the equilibrium properties of these highly anisotropic QDs.

We have also explored the dynamical properties of the QDs through
head-on collisions between identical droplets along different
directions and through their rotational dynamics under reorientation
of the microwave-controlled DDI. The collision outcomes exhibit a
pronounced directional dependence. Collisions along the $x$ axis lead
to either quasi-elastic rebound or merger, whereas collisions along the
 $y$ and $z$ axes are destructive and result in fragmentation without
 a long-lived self-bound remnant. In addition, direct simulations of
 the rotational dynamics show that a self-bound QD can dynamically
 follow the reorientation of the non-axisymmetric DDI, undergoing
 orientation-dependent deformation while maintaining its spatial
 localization. These results demonstrate that non-axisymmetric
 microwave-dressed DDIs provide versatile control over the
 equilibrium structure, collisional behavior, and
 three-dimensional rotational dynamics of self-bound quantum droplets.

As an extension of the present work, it would be interesting to investigate
collective excitations~\cite{Wachtler2016,Chomaz2018,FerrierBarbut2018} of
non-axisymmetric QDs,
because the pronounced anisotropic
deformation, found here, breaks rotational symmetry and may split and mix
low-lying shape modes. It would also be important to clarify specific
mechanisms which account for outcomes of the direction-dependent collisions,
reported in this work.
Actually, the different outcomes suggest action of direction-dependent
coupling between translational motion, internal deformation, and fragmentation.
A systematic mode-resolved analysis of the corresponding energy-transfer
pathways would require a dedicated dynamical study
(cf. Refs.~\cite{Baillie2016,Ferioli2019,Adhikari2017}),
which is a subject for further work.

\section{ACKNOWLEDGMENTS}

We appreciate valuable discussions with Prof. D. Baillie (University of
Otago, New Zealand). This work was supported by NNSFC (China) through Grants
No. 12274077 and 12475014, Guangdong Basic and Applied Basic Research
Foundation through grants No. 2024A1515030131, 2025A1515011128,
2023A1515110198, and 2023A1515010770, the Research Fund of Guangdong-Hong
Kong-Macao Joint Laboratory for Intelligent Micro-Nano Optoelectronic
Technology, through grant No. 2020B1212030010.

\appendix

\section{The derivation of $\mathcal{Q}_{5}(\protect\epsilon_{dd}^{(0)},%
\protect\epsilon_{dd}^{(2)})$}

The calculation of $\mathcal{Q}_{5}(\epsilon _{dd}^{(0)},\epsilon
_{dd}^{(2)})$ is derived by D. Baillie in Ref. \cite{Baillie}. In this
appendix, we briefly reproduce the derivation, to facilitate understanding
of properties of this function.

By introducing a spherical cutoff of radius $R_{c}$, the Fourier transform
of the dipole-dipole interaction given in Eq.~(\ref{eq:Fourier_DDI}) can be rewritten as
\begin{equation}
\begin{aligned}
{\tilde{U}_{\mathrm{dd}}^{(R_{c})}}(\mathbf{k})
&=\tilde{U}_{\mathrm{dd}}(\mathbf{k})s(kR_c)\\
&= G_s\Bigl[
\epsilon_{dd}^{(0)}\left(3\cos^{2}\theta_{k}-1\right)\\
&\hphantom{\mathrel{=}} -\sqrt{3}\epsilon_{dd}^{(2)}
\sin^{2}\theta_{k}\cos 2\varphi_{k}
\Bigr]s(kR_c),
\end{aligned}
\end{equation}

where $\theta_k$ and $\varphi_k$ denote the polar and azimuthal angles of
the wave vector $\mathbf{k}$, respectively, and the cutoff function $s(kR_c)$ is
\begin{equation}
\begin{gathered} s(kR_c)=1+3(kR_c)^{-2}\cos(kR_c)-3(kR_c)^{-3}\sin(kR_c).
\end{gathered}
\end{equation}%

In the limit of $R_{c}\rightarrow \infty $, one has $s(kR_{c})\rightarrow 1$%
, so that $\tilde{U}_{\mathrm{dd}}^{R_{c}}(\mathbf{k})\rightarrow \tilde{U}_{%
\mathrm{dd}}(\mathbf{k})$.

To evaluate the angular average, the integrations over $\theta _{k}$ can be
carried out analytically, yielding
\begin{equation}
\begin{gathered}
\mathcal{Q}_5(x,0)=\frac{5(x-1)^3}{16\sqrt{3x}}\Bigl[\ln\bigl(\sqrt{1+2x}-%
\sqrt{3x}\bigr)-\frac{\ln(1-x)}{2}\Bigr]\\+\frac{1}{16}(11+4x+9x^2)%
\sqrt{1+2x}, \end{gathered}
\end{equation}%

The remaining dependence on $\phi _{k}$ is represented by an auxiliary
function,
\begin{gather}
I(\phi _{k})=\left( 1-\frac{2}{\sqrt{3}}\epsilon _{dd}^{(2)}\cos 2\phi
_{k}\right) ^{5/2}  \notag \\
\times \mathcal{Q}_{5}\!\left( \frac{\sqrt{3}\epsilon _{dd}^{(0)}+\epsilon
_{dd}^{(2)}\cos 2\phi _{k}}{\sqrt{3}-2\epsilon _{dd}^{(2)}\cos 2\phi _{k}}%
,0\right) .
\end{gather}%

The full coefficient $\mathcal{Q}_{5}(\epsilon _{dd}^{(0)},\epsilon
_{dd}^{(2)})$ is then obtained as
\begin{equation}
\text{Re}\{\mathcal{Q}_{5}(\epsilon _{dd}^{(0)},\epsilon _{dd}^{(2)})\}=%
\frac{2}{\pi }\int_{0}^{\pi /2}d\phi _{k}\,|\text{Re}\{I(\phi _{k})\}|,
\end{equation}%
\begin{equation}
\text{Im}\{\mathcal{Q}_{5}(\epsilon _{dd}^{(0)},\epsilon _{dd}^{(2)})\}=%
\frac{2}{\pi }\int_{0}^{\pi /2}d\phi _{k}\,|\text{Im}\{I(\phi _{k})\}|,
\end{equation}%

In parameter regions where $\mathcal{Q}_{5}$ is complex, only its real part
is retained in the extended GPE. The so obtained values of Re[$\mathcal{Q}%
_{5}$] for different values of $\epsilon _{dd}^{(2)}$ and $\epsilon
_{dd}^{(0)}=0$ are collected in Table I.

\begin{table}[htbp]
\caption{Values of Re$[\mathcal{Q}_5]$ for selected $\protect\epsilon%
^{(2)}_{dd} $ (with $\protect\epsilon^{(0)}_{dd}=0$).}
\label{tab:correspondence}\centering
\setlength{\heavyrulewidth}{0.4pt} %
\setlength{\lightrulewidth}{0.4pt} 
\setlength{\tabcolsep}{48pt} %
\vspace{4pt} %
\par
\vspace{4pt} %
\par
\hrule height 0.4pt %
\vspace{2pt} %
\hrule height 0.4pt %
\par
\vspace{4pt} %
\setlength{\aboverulesep}{3pt} \renewcommand{\arraystretch}{1.1}
\begin{tabular}{cc}
\bfseries $\epsilon^{(2)}_{dd}$ & \bfseries Re$[\mathcal{Q}_5]$ \\
\midrule 1 & 2.43 \\
2 & 6.61 \\
3 & 13.22 \\
4 & 23.29 \\
5 & 37.02 \\
\end{tabular}
\vspace{3pt} %
\hrule height 0.4pt \vspace{2pt} %
\hrule height 0.4pt
\end{table}


\begin{thebibliography}{99}
\bibitem{Krem2008} R. V. Krems, Cold controlled chemistry, Phys. Chem. Chem.
Phys. \textbf{10}, 4079 (2008).

\bibitem{Baranov2008} M. Baranov, Theoretical progress in many-body physics
with ultracold dipolar gases, Phys. Rep. \textbf{464}, 71 (2008).

\bibitem{Ni2008} K.-K. Ni, S. Ospelkaus, M. H. G. De Miranda, A. Pe'er, B.
Neyenhuis, J. J. Zirbel, S. Kotochigova, P. S. Julienne, D. S. Jin, and J.
Ye, A High Phase-Space-Density Gas of Polar Molecules, Science \textbf{322},
231 (2008).

\bibitem{Moses2017} S. A. Moses, J. P. Covey, M. T. Miecnikowski, D. S. Jin,
and J. Ye, New frontiers with quantum gases of polar molecules, Nat. Phys.
\textbf{13}, 13 (2017).

\bibitem{Lahaye2009} T. Lahaye, C. Menotti, L. Santos, M. Lewenstein, and T.
Pfau, The physics of dipolar bosonic quantum gases, Rep. Prog. Phys. \textbf{%
72}, 126401 (2009).

\bibitem{Baranov2012} M. A. Baranov, M. Dalmonte, G. Pupillo, and P. Zoller,
Condensed matter theory of dipolar quantum gases, Chem. Rev. \textbf{112},
5012 (2012).

\bibitem{Micheli2006} A. Micheli, G. K. Brennen, and P. Zoller, A toolbox
for lattice spin models with polar molecules, Nat. Phys. \textbf{2}, 341
(2006).

\bibitem{Carr2009} L. D. Carr, D. DeMille, R. V. Krems, and J. Ye, Cold and
ultracold molecules: science, technology, and applications, New J. Phys.
\textbf{11}, 055049 (2009).

\bibitem{Altman2021} E. Altman, K. R. Brown, G. Carleo, L. D. Carr, E.
Demler, C. Chin, B. DeMarco, S. E. Economou, M. A. Eriksson, K.-M. C. Fu, et
al., Quantum simulators: Architectures and opportunities, PRX Quantum
\textbf{2}, 017003 (2021).


\bibitem{Baillie2002} D. DeMille, Quantum computation with trapped polar
molecules, Phys. Rev. Lett. \textbf{88}, 067901 (2002).

\bibitem{Rabl2006} P. Rabl, D. DeMille, J. M. Doyle, M. D. Lukin, R. J.
Schoelkopf, and P. Zoller, Hybrid quantum processors: molecular ensembles as
quantum memory for solid state circuits, Phys. Rev. Lett. \textbf{97},
033003 (2006).

\bibitem{Ospelkaus2010} K.-K. Ni, S. Ospelkaus, D. Wang, G. Qu\'{e}m\'{e}%
ner, B. Neyenhuis, M. H. G. De Miranda, J. L. Bohn, J. Ye, and D. S. Jin,
Dipolar collisions of polar molecules in the quantum regime, Nature \textbf{%
464}, 1324 (2010).

\bibitem{Quemener2012} G. Qu\'{e}m\'{e}ner and P. S. Julienne, Ultracold
molecules under control!, Chem. Rev. \textbf{112}, 4949 (2012).

\bibitem{Bohn2017} J. L. Bohn, A. M. Rey, and J. Ye, Cold molecules:
Progress in quantum engineering of chemistry and quantum matter, Science
\textbf{357}, 1002 (2017).

\bibitem{XY2018} X. Ye, M. Guo, M. L. Gonz\'{a}lez-Mart\'{\i}nez, G. Qu\'{e}m%
\'{e}ner, and D. Wang, Collisions of ultracold $^{23}$Na$^{87}$Rb molecules
with controlled chemical reactivities, Sci. Adv. \textbf{4}, eaaq0083 (2018).

\bibitem{Bause2023} R. Bause, A. Christianen, A. Schindewolf, I. Bloch, and
X.-Y. Luo, Ultracold sticky collisions: theoretical and experimental status,
J. Phys. Chem. A \textbf{127}, 729 (2023).

\bibitem{Gorshkov2008} A. V. Gorshkov, P. Rabl, G. Pupillo, A. Micheli,
P. Zoller, M. D. Lukin, and H. P. Büchler, Repulsive shield between polar molecules,
Phys. Rev. Lett. \textbf{101}, 73201 (2008).

\bibitem{Karman2018} T. Karman and J. M. Hutson, Microwave shielding of
ultracold polar molecules, Phys. Rev. Lett. \textbf{121}, 163401 (2018).

\bibitem{Lassabliere2018} L. Lassabli\`{e}re and G. Qu\'{e}m\'{e}ner,
Controlling the scattering length of ultracold dipolar molecules, Phys. Rev.
Lett. \textbf{121}, 163402 (2018).

\bibitem{Anderegg2021} L. Anderegg, S. Burchesky, Y. Bao, S. S. Yu, T.
Karman, E. Chae, K.-K. Ni, W. Ketterle, and J. M. Doyle, Observation of
microwave shielding of ultracold molecules, Science \textbf{373}, 779 (2021).

\bibitem{Schindewolf2022} A. Schindewolf, R. Bause, X.-Y. Chen, M. Duda, T.
Karman, I. Bloch, and X.-Y. Luo, Evaporation of microwave-shielded polar
molecules to quantum degeneracy, Nature \textbf{607}, 677 (2022).

\bibitem{Chen2023} X.-Y. Chen, A. Schindewolf, S. Eppelt, R. Bause, M. Duda,
S. Biswas, T. Karman, T. Hilker, I. Bloch, and X.-Y. Luo, Field-linked
resonances of polar molecules, Nature \textbf{614}, 59 (2023).

\bibitem{Lin2023} J. Lin, G. Chen, M. Jin, Z. Shi, F. Deng, W. Zhang, G. Qu%
\'{e}m\'{e}ner, T. Shi, S. Yi, and D. Wang, Microwave shielding of bosonic
NaRb molecules, Phys. Rev. X \textbf{13}, 031032 (2023).

\bibitem{Bigagli2023} N. Bigagli, C. Warner, W. Yuan, S. Zhang, I.
Stevenson, T. Karman, and S. Will, Collisionally stable gas of bosonic
dipolar ground state molecules, Nat. Phys. \textbf{19}, 1579 (2023).

\bibitem{Deng2023} F. Deng, X.-Y. Chen, X.-Y. Luo, W. Zhang, S. Yi, and T.
Shi, Effective potential and superfluidity of microwave-shielded polar
molecules, Phys. Rev. Lett. \textbf{130}, 183001 (2023).

\bibitem{Chen2024} X.-Y. Chen, S. Biswas, S. Eppelt, A. Schindewolf, F.
Deng, T. Shi, S. Yi, T. A. Hilker, I. Bloch, and X.-Y. Luo, Ultracold
field-linked tetratomic molecules, Nature \textbf{626}, 283 (2024).

\bibitem{Dutta2025} J. Dutta, Universality in the microwave shielding of
ultracold polar molecules, Phys. Rev. Res. \textbf{7}, 023164 (2025).

\bibitem{Karman2025} T. Karman, N. Bigagli, W. Yuan, S. Zhang, I. Stevenson,
and S. Will, Double microwave shielding, PRX Quantum \textbf{6}, 020358
(2025).

\bibitem{Bigagli2024} N. Bigagli, W. Yuan, S. Zhang, B. Bulatovic, T.
Karman, I. Stevenson, and S. Will, Observation of Bose-Einstein condensation
of dipolar molecules, Nature \textbf{631}, 289 (2024).

\bibitem{Yuan2025} W. Yuan, S. Zhang, N. Bigagli, H. Kwak, C. Warner, T.
Karman, I. Stevenson, and S. Will, Extreme loss suppression and wide
tunability of dipolar interactions in an ultracold molecular gas,
arXiv:2505.08773.

\bibitem{Shi2025} Z. Shi, Z. Huang, F. Deng, W.-J. Jin, S. Yi, T. Shi, and
D. Wang, Bose-Einstein condensate of ultracold sodium-rubidium molecules
with tunable dipolar interactions, Nat. Phys. (2026).

\bibitem{Deng2025} F. Deng, X. Hu, W.-J. Jin, S. Yi, and T. Shi,
Two- and many-body physics of ultracold molecules dressed by dual microwave fields,
Nat. Commun. \textbf{16}, 11219 (2025).
\bibitem{Arnone2026} T. Arnone Cardinale, T. Bland, and S. M. Reimann,
Exploring supersolids of single-microwave shielded molecules via exact
and mean-field theories, Commun. Phys. \textbf{9}, 191 (2026).
\bibitem{Jin2025} W.-J. Jin, F. Deng, S. Yi, and T. Shi, Bose-einstein condensates of
microwave-shielded polar molecules, Phys. Rev. Lett. \textbf{134}, 233003 (2025).

\bibitem{Li2024} G. Li, X. Jiang, B. Liu, Z. Chen, B. A. Malomed, and Y. Li,
Two-dimensional anisotropic vortex quantum droplets in dipolar Bose-Einstein
condensates, Front. Phys. \textbf{19}, 22202 (2024).


\bibitem{FerrierBarbut2016} I. Ferrier-Barbut, H. Kadau, M. Schmitt, M.
Wenzel, and T. Pfau, Observation of quantum droplets in a strongly dipolar
bose gas, Phys. Rev. Lett. \textbf{116}, 215301 (2016).

\bibitem{Schmitt2016} M. Schmitt, M. Wenzel, F. B\"{o}ttcher, I.
Ferrier-Barbut, and T. Pfau, Self-bound droplets of a dilute magnetic
quantum liquid, Nature \textbf{539}, 259 (2016).

\bibitem{Kadau2016} H. Kadau, M. Schmitt, M. Wenzel, C. Wink, T. Maier, I.
Ferrier-Barbut, and T. Pfau, Observing the Rosensweig instability of a
quantum ferrofluid, Nature \textbf{530}, 194 (2016).

\bibitem{Boudjemaa2018} A. Boudjem\^{a}a, Fluctuations and quantum
self-bound droplets in a dipolar bose-bose mixture, Phys. Rev. A \textbf{98}%
, 033612 (2018).

\bibitem{Semeghini2018} G. Semeghini, G. Ferioli, L. Masi, C. Mazzinghi, L.
Wolswijk, F. Minardi, M. Modugno, G. Modugno, M. Inguscio, and M. Fattori,
Self-bound quantum droplets of atomic mixtures in free space, Phys. Rev.
Lett. \textbf{120}, 235301 (2018).

\bibitem{Cabrera2018} C. R. Cabrera, L. Tanzi, J. Sanz, B. Naylor, P.
Thomas, P. Cheiney, and L. Tarruell, Quantum liquid droplets in a mixture of
Bose-Einstein condensates, Science \textbf{359}, 301 (2018).

\bibitem{Chomaz2022} L. Chomaz, I. Ferrier-Barbut, F. Ferlaino, B.
Laburthe-Tolra, B. L. Lev, and T. Pfau, Dipolar physics: a review of
experiments with magnetic quantum gases, Rep. Prog. Phys. \textbf{86},
026401 (2022).
\bibitem{Xu2025}P. Xu and G. Chen, Effective potential and scattering
length of shielded polar molecules, Phys. Rev. A \textbf{112}, 043306 (2025).
\bibitem{Ho2026}C. J. Ho, J. Dutta, B. Mukherjee, J. M. Hutson,
and M. R. Tarbutt, Tuning interactions between static-field-shielded
polar molecules with microwaves, Phys. Rev. Research \textbf{8}, 023087 (2026).

\bibitem{experiment} S. Zhang, W. Yuan, N. Bigagli, H. Kwak, T. Karman, I.
Stevenson, and S. Will, Observation of self-bound droplets of ultracold
dipolar molecules, Nature \textbf{651}, 601 (2026).
\bibitem{Langen2025} T. Langen, J. Boronat, J. Sánchez-Baena, R. Bombín, T. Karman, and F. Mazzanti,
Dipolar droplets of strongly interacting molecules, Phys. Rev. Lett. \textbf{134}, 53001 (2025).
\bibitem{langen2025} T. Langen, J. Boronat, J. S\'{a}nchez-Baena,
 R. Bomb\'{\i}n, T. Karman, and F. Mazzanti, Dipolar Droplets of
 Strongly Interacting Molecules, Phys. Rev. Lett. \textbf{134}, 053001 (2025).
\bibitem{Schmidt2022} M. Schmidt, L. Lassabli\`{e}re, G. Qu\'{e}m\'{e}ner,
 and T. Langen, Self-bound dipolar droplets and supersolids in molecular
 Bose-Einstein condensates, Phys. Rev. Research \textbf{4}, 013235 (2022).

\bibitem{Petrov2015} D. S. Petrov, Quantum Mechanical Stabilization of a
Collapsing Bose-Bose Mixture, Phys. Rev. Lett. \textbf{115}, 155302 (2015).

\bibitem{Petrov2016} D. S. Petrov and G. E. Astrakharchik, Ultradilute
Low-Dimensional Liquids, Phys. Rev. Lett. \textbf{117}, 100401 (2016).

\bibitem{Bisset2016} R. N. Bisset, R. M. Wilson, D. Baillie, and P. B.
Blakie, Ground-state phase diagram of a dipolar condensate with quantum
fluctuations, Phys. Rev. A \textbf{94}, 033619 (2016).

\bibitem{Lima2011} A. R. P. Lima and A. Pelster, Quantum fluctuations in
dipolar bose gases, Phys. Rev. A \textbf{84}, 041604 (2011).

\bibitem{FerrierBarbut2016Review} I. Ferrier-Barbut, M. Schmitt, M. Wenzel,
H. Kadau, and T. Pfau, Liquid quantum droplets of ultracold magnetic atoms,
J. Phys. B: At. Mol. Opt. Phys. \textbf{49}, 214004 (2016).

\bibitem{Guo2020} M. Guo and T. Pfau, A new state of matter of quantum
 droplets, Front. Phys. \textbf{16}, 32202 (2020).


\bibitem{Baillie} D. Baillie, Symmetry and self-bound droplets in dipolar
molecular gases, Phys. Rev. Res. \textbf{8}, 023219 (2026).

\bibitem{Zhang2026} W. Zhang, H. Liu, F. Deng, K. Chen, S. Yi, and T. Shi,
 Supersolid Phases in Ultracold Gases of Microwave Shielded Polar Molecules, arXiv:2506.23820.
\bibitem{Schindewolf2026} A. Schindewolf, J. Hertkorn, I. Stevenson, M. Ciardi,
P. Groß, D. Wang, T. Karman, G. Quéméner, S. Will, T. Pohl, and T. Langen, Colloquium:
Strongly dipolar molecular Bose-Einstein condensates: From few- to many-body
physics, Rev. Mod. Phys. (2026).
\bibitem{Zhao2026}Z. Zhao, G. Li, Z. Chen, H.-B. Luo, B. Liu, B. A. Malomed, and Y. Li,
 Bulk modulus of three-dimensional quantum droplets, Phys. Rev. A \textbf{113}, 43315 (2026).

\bibitem{ITP} J. Yang, Nonlinear Waves in Integrable and Non-Integrable
Systems (SIAM, Philadelphia, 2010).


\bibitem{MATLAB_fminsearch}
The MathWorks, Inc., \textit{fminsearch}, MATLAB Documentation,
Natick, MA, USA.

\bibitem{VK} N. G. Vakhitov and A. A. Kolokolov, Stationary solutions of the
wave equation in a medium with nonlinearity saturation, Radiophys. Quantum
Electron. \textbf{16}, 783 (1973).

\bibitem{Singh2026} S. Singh and C. H. Greene, Efimov effect in ultracold
 microwave-shielded polar molecules, Phys. Rev. Lett. \textbf{136}, 233402 (2026).

\bibitem{Wachtler2016} F. W\"{a}chtler and L. Santos, Ground-state
properties and elementary excitations of quantum droplets in dipolar
bose-einstein condensates, Phys. Rev. A \textbf{94}, 043618 (2016).

\bibitem{Chomaz2018} L. Chomaz, R. M. W. Van Bijnen, D. Petter, G. Faraoni,
S. Baier, J. H. Becher, M. J. Mark, F. W\"{a}chtler, L. Santos, and F.
Ferlaino, Observation of roton mode population in a dipolar quantum gas,
Nat. Phys. \textbf{14}, 442 (2018).

\bibitem{FerrierBarbut2018} I. Ferrier-Barbut, M. Wenzel, F. B\"{o}ttcher,
T. Langen, M. Isoard, S. Stringari, and T. Pfau, Scissors mode of dipolar
quantum droplets of dysprosium atoms, Phys. Rev. Lett. \textbf{120}, 160402
(2018).

\bibitem{Baillie2016} D. Baillie, R. M. Wilson, R. N. Bisset, and P. B.
Blakie, Self-bound dipolar droplet: a localized matter-wave in free space,
Phys. Rev. A \textbf{94}, 021602(R) (2016).

\bibitem{Ferioli2019} G. Ferioli, G. Semeghini, L. Masi, G. Giusti, G.
Modugno, M. Inguscio, A. Gallem\'{\i}, A. Recati, and M. Fattori, Collisions
of Self-Bound Quantum Droplets, Phys. Rev. Lett. \textbf{122}, 090401 (2019).

\bibitem{Adhikari2017} S. K. Adhikari, Statics and dynamics of a self-bound
matter-wave quantum ball, Phys. Rev. A \textbf{95}, 023606 (2017).
\end{thebibliography}
\end{document}